\documentclass[12pt, a4paper]{article}

\usepackage{amsmath, amssymb, amsthm}  
\usepackage{geometry}    
\usepackage{float}    
\usepackage{mathtools}
\usepackage{graphicx}                 
\usepackage{hyperref}                  
\usepackage[svgnames]{xcolor}           
\usepackage{mathrsfs}
\usepackage{cite}
\usepackage{bm}

\usepackage{tikz}                     
\usetikzlibrary{calc, shapes, arrows.meta, positioning, decorations.pathmorphing, patterns}
\usepackage{fancyhdr}                  
\usepackage{titlesec}                   
\usepackage{enumitem}

\titleformat{\section}{\Large\bfseries\color{blue!40!black}}{}{0em}{}[\titlerule]
\titleformat{\subsection}{\large\bfseries\color{blue}}{}{0em}{}
\titleformat{\subsubsection}{\bfseries\color{purple}}{}{0em}{}

\setlist[itemize]{leftmargin=*, itemsep=2pt, topsep=5pt}

\usepackage{geometry}
\definecolor{lime}{HTML}{A6CE39}
\newcommand{\orcidicon}{%
	\begin{tikzpicture}
	\draw[lime, fill=lime] (0,0)
	circle [radius=0.16]
	node[white] {{\fontfamily{qag}\selectfont \tiny ID}};
	\draw[white, fill=white] (-0.0625,0.095)
	circle [radius=0.007];
	\end{tikzpicture}   \hspace{-2mm}
}
\newcommand\orcidFaical{{\href{https://orcid.org/0000-0002-2977-0821}{\orcidicon}}}
\begin{document}
\vspace{1cm}

\begin{titlepage}
    \centering
    \vspace{1cm}

    {\Large \bfseries \textcolor{Black}{Celestial Holography, A Hitchhiker's Guide to the\\[0.2cm] Celestial Sphere}} \\[0.4cm]
    {\large \textit{Or, How I Learned to Stop Worrying and Love the Leaky Boundary}} \\[2cm]

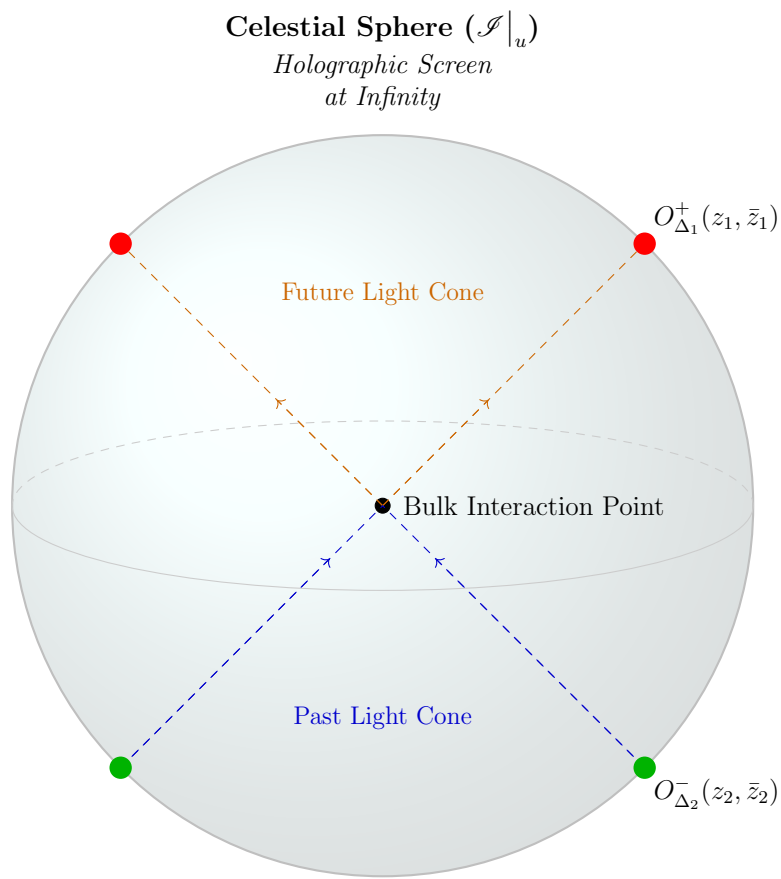
\begin{figure}[H]
\hspace{3.1cm}
\begin{tikzpicture}[line join=round, line cap=round,scale=1.4]

    \coordinate (Bulk) at (0,0); 
    \def\R{3.5} 

    \shade[ball color=cyan!20, opacity=0.2] (Bulk) circle (\R);
    
    \draw[thick, color=gray!50] (Bulk) circle (\R);
    
    \draw[dashed, color=gray!40] (\R,0) arc (0:180:\R\space and 0.8);
    \draw[color=gray!40] (\R,0) arc (0:-180:\R\space and 0.8);

    \filldraw[black] (Bulk) circle (2pt) node[ xshift=2cm, scale=0.85] {Bulk Interaction Point};

    \def\angA{45}
    \def\angB{135}

    \draw[dashed, orange!80!black, ->] (Bulk) -- (\angA:\R);
     \draw[dashed, orange!80!black, ->] (Bulk)--(1,1) ;
    \draw[dashed, orange!80!black, ->] (Bulk) -- (\angB:\R);
    \draw[dashed, orange!80!black, ->] (Bulk)--(-1,1);
    \fill[red] (\angA:\R) circle (3pt) node[above right, black, scale=0.85] {$O^+_{\Delta_1}(z_1, \bar{z}_1)$};
    \fill[red] (\angB:\R) circle (3pt);
    \node[orange!80!black, scale=0.8] at (0, 2) {Future Light Cone};

    \def\angC{-45}
    \def\angD{-135}
    
    \draw[dashed, blue!80!black, <-] (\angC:\R) -- (Bulk);
    \draw[dashed, blue!80!black, ->] (\angC:\R) -- (0.5,-0.5);
    \draw[dashed, blue!80!black, <-] (\angD:\R) -- (Bulk);
    \draw[dashed, blue!80!black, ->] (\angD:\R) -- (-0.5,-0.5);
    \fill[green!70!black] (\angC:\R) circle (3pt) node[below right, black, scale=0.85] {$O^-_{\Delta_2}(z_2, \bar{z}_2)$};
    \fill[green!70!black] (\angD:\R) circle (3pt);
    \node[blue!80!black, scale=0.8] at (0, -2) {Past Light Cone};

    \node[scale=0.9, font=\bfseries] at (0, \R+1) {Celestial Sphere ($\mathscr{I}\big|_u$)};
    \node[scale=0.85, align=center] at (0, \R+0.5) {\textit{Holographic Screen} \\\textit{at Infinity}};

\end{tikzpicture}
\end{figure}
    \vspace{1cm}
    \hspace{1cm}
    \begin{center}
            {\large  Faïçal Barzi\orcidFaical\!\!} \\
    {\small LPTHE, Physics Department, Faculty of Sciences, Ibnou Zohr University, Agadir, Morocco. \\
CRMEF, Regional Center for Education and Training Professions Marrakesh, Morocco.}\\
{\small faical.barzi@edu.uiz.ac.ma}\\[0.3cm]
    {\large  \today} \\
    \vfill
        \end{center}
\end{titlepage}

 \begin{minipage}{0.9\textwidth}
        \begin{center}
            \textbf{Abstract} 
        \end{center}
        \paragraph{}This primer offers an first introduction to Celestial Holography, a rapidly evolving framework aiming to describe quantum gravity in asymptotically flat spacetimes. We start from the established foundation of the AdS/CFT correspondence \cite{Maldacena1999} before venturing into the wilds of Minkowski space. The journey covers the essential Bondi-Sachs formalism \cite{Madler2016} and the geometry of null infinity ($\mathscr{I}^\pm$), the discovery of the infinite-dimensional BMS asymptotic symmetry group. We explain the core holographic map given by Mellin transform that recasts four-dimensional scattering amplitudes as correlation functions of a conjectured 2D Celestial Conformal Field Theory (CCFT) on the celestial sphere. The presentation is made intuitive but the mathematics is still precise (hopefully!). It is designed to equip newcomers with both the conceptual landscape and the technical vocabulary to explore this exciting frontier.
    \end{minipage}

\tableofcontents

\section{The Lay of the Land: Why Celestial Holography?}

Our story begins not with an answer, but with a location problem. For over two decades, theoretical physicists have wielded a tool of breathtaking power for studying quantum gravity: the Anti-de Sitter/Conformal Field Theory (AdS/CFT) correspondence, first concretely proposed by Juan Maldacena in 1997 \cite{Maldacena1999}. This \textit{holographic} duality states that a theory of gravity in a $(d+1)$-dimensional spacetime with negative curvature called an Anti-de Sitter or AdS space is completely equivalent to a non-gravitational quantum field theory. Specifically, a Conformal Field Theory (CFT) living on its $d$-dimensional \textit{timelike} boundary.\\

It's a theoretical physicist's dream: intractable quantum gravity problems in the bulk can be translated into (still difficult, but more familiar) problems in a well-defined quantum field theory on the boundary. AdS/CFT has been a roaring success, providing deep insights into quantum gravity, black holes, and even condensed matter systems.\\

There's just one snag, and it's a cosmological one: \textit{our universe is not asymptotically AdS}. Observations strongly suggest that on large scales, our universe is approximately flat (\textit{Minkowski space}) or has a very small positive cosmological constant, a so-called \textit{de Sitter space}. We are, to borrow a phrase, \textit{living in the wrong holographic universe for the AdS/CFT toolkit}. This realization is the primary engine driving celestial holography program\cite{Donnay2022,Donnay2023,Pasterski2021,Raclariu2021}, that is, the urgent desire to formulate a \textbf{holographic principle tailored to our own, asymptotically flat reality}.\\

\begin{figure}[H]
\label{fig:inception}
    \centering
    \begin{tikzpicture}[scale=1.3]
        \draw[thick, ->] (0,0) -- (10,0);
        \foreach \x/\year in {1/1960s, 4/1997, 7/2010s, 9.5/Today}
            \draw (\x,0.1) -- (\x,-0.1) node[below] {\year};
        \node[align=center, font=\footnotesize] at (1, 0.8) {Bondi, Sachs,\\Metzner\\BMS Symmetry};
        \node[align=center, font=\footnotesize] at (4, 1.2) {Maldacena\\AdS/CFT};
        \node[align=center, font=\footnotesize] at (7, 0.8) {Strominger et al.\\BMS $\leftrightarrow$ Soft Theorems\\Memory, Soft Hair};
        \node[align=center, font=\footnotesize] at (9.5, 1.2) {Celestial\\Holography\\Program};
        \foreach \x in {1,4,7,9.5}
            \draw[dashed, gray] (\x,0.6) -- (\x,0.1);
    \end{tikzpicture}
    \caption{\textbf{The conceptual evolution} leading to modern celestial holography, weaving together asymptotic symmetries from the 1960s, the holographic principle from the 1990s, and their physical unification in the 2010s.}
\end{figure}
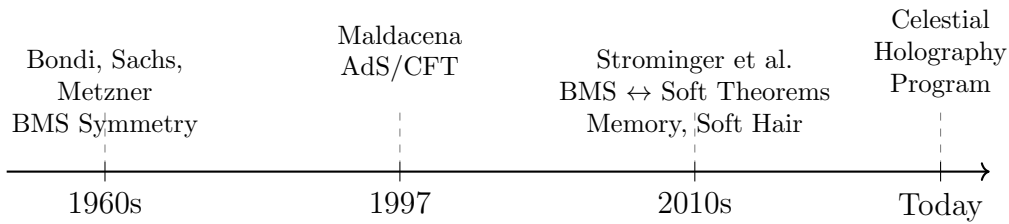

But if AdS/CFT is the well-mapped continent, flat space holography has long been a mysterious wilderness. The journey into this wilderness didn't start yesterday (see Fig.\ref{fig:inception}). Its trails were first blazed in the 1960s by Hermann Bondi, Rainer Sachs, and others \cite{Madler2016}. They developed the Bondi-Sachs formalism to rigorously describe isolated gravitational systems (like stars and black holes) emitting radiation. In doing so, they identified the true asymptotic symmetry group of flat spacetime and to the surprise of everybody it turns up not to be the famous finite 10-dimensional Poincaré group, but a vast, \textit{infinite-dimensional} group we now call the BMS for Bondi-Metzner-Sachs\footnote{Technically its the  BvBMS  for Bondi-van der Burg-Metzner-Sachs, but the "v" being a logic operator was inadvertently simplified!} group, which includes so-called \textit{supertranslations} and \textit{superrotations}.\\'

For decades, these BMS symmetries were considered a mathematical curiosity, an artifact of the formalism with no physical consequences or even a nuisance. That changed dramatically in the 2010s, led by work of Andrew Strominger and colleagues. They made a stunning connection, they showed that the Ward identities associated with BMS supertranslations are mathematically \textit{identical} to Weinberg's soft graviton theorem \cite{Strominger2017}. This theorem, a cornerstone of quantum field theory, dictates that any scattering process must emit a characteristic cloud of very low-energy \textbf{soft} gravitons. Suddenly, BMS symmetries were not mere redundancies but physical symmetries with observable consequences, governing the infrared structure of spacetime.\\

This revelation opened the floodgates. It led to the understanding of \textbf{gravitational memory}, that is, the permanent displacement of test masses by passing gravitational waves, as a direct imprint of BMS symmetry \cite{Strominger2017}. It sparked the provocative \textit{soft hair} proposal for black holes, suggesting that an infinite number of BMS charges on the horizon could label microstates and address the information paradox \cite{Hawking2015}. Most importantly for our story, it provided the crucial clue about the holographic dual for flat spacetime; it must be built on this infinite-dimensional asymptotic symmetry structure.\\

Thus, celestial holography was born from a convergence of the mature, \textit{top-down} architecture of AdS/CFT, the \textit{bottom-up}, physics-first approach of asymptotic symmetries, and the tantalizing hints of a new duality encoded in soft theorems. It asks \textbf{The question}: \textit{Can we use the BMS group as our guide to construct a holographic dictionary where the sky at infinity, called the \textbf{celestial sphere}, becomes the screen for the hologram?}

The goal is ambitious since one tries to describe four-dimensional quantum gravity processes, like the scattering of particles or the evaporation of black holes, using a potentially simpler, lower-dimensional theory living on the boundary of Minkowski space.\\

\begin{figure}[H]
    \label{fig:ry}
    \centering
\begin{tikzpicture}[line join=round,scale=1.29]
    \begin{scope}[shift={(-4,0)}]
        \node at (0,2.8) {\textbf{Minkowski (Flat) Spacetime}};
        \node[scale=0.8] at (0,-2.8) {Leaky Null Boundary};

        \draw[thick,blue] (0,-2) -- (2,0) node[right] {$i^0$} -- (0,2);
        \draw[thick,blue] (0,-2) node[below] {$i^-$} -- (-2,0) -- (0,2) node[above] {$i^+$};
        
        \node[rotate=45,blue] at (1.2,1.2) {$\mathscr{I}^+$};
        \node[rotate=-45,blue] at (1.2,-1.2) {$\mathscr{I}^-$};
        
        \foreach \y in {0.5, 1.5,2} {
            \draw[-{Stealth}, dashed,red] (0, \y-1) -- (1, \y);
        };
      \foreach \z in {0.5, 1.5} {
            \draw[ -{Stealth}, dashed,red] (0+2, \z-1-2) -- (-1+2, \z-2);
        };
        
        \node[ scale=0.9,red] at (2.4, 1.2) {Radiation};
        \node[ scale=0.9,red] at (2.4, 0.9) {Leaking Out};
    \end{scope}

    \begin{scope}[shift={(3.5,0)}]
        \node at (0,2.8) {\textbf{Anti-de Sitter (AdS) Spacetime}};
        \node[scale=0.8] at (0,-2.8) {Timelike Reflective Boundary};

         \draw[ultra thick,blue] (-1.5,-2) -- (-1.5,2);
        \draw[ultra thick,blue] (1.5,-2) -- (1.5,2);
        % \node[right, scale=0.9] at (1.5, 0) {Radiation};
        % \node[right, scale=0.9] at (1.5, -0.3) {Reflected back};
        
        \draw[-{Stealth}, dashed] (-0.05, -0.5-1.1) -- (1.5-0.05, 0.5-0.3);
        \draw[ -{Stealth}, dashed] (1.5, 0.5-0.3) -- (-0., 1.5+0.4);
        
        \draw[-{Stealth}, dashed] (-0.05, -0.5-1.1-0.8) -- (1.5-0.05, 0.5-0.3-0.8);
        \draw[ -{Stealth}, dashed] (1.5, 0.5-0.3-0.8) -- (-0., 1.5+0.4-0.8);
         \node[scale=0.9] at (-0.2, 0.2) {Radiation};
         \node[scale=0.9] at (-0.2, -0.1) {Reflected Back};
    \end{scope}

\end{tikzpicture}
\caption{\small \textbf{The Boundary Problem.} The AdS/CFT correspondence relies on a reflective, timelike boundary. Celestial holography must contend with the leaky, null boundary of flat spacetime, $\mathscr{I}^+$, where radiation freely escapes. This fundamental difference is the source of both the challenge and the rich new structure (like BMS symmetry) of flat space holography.}
\end{figure}
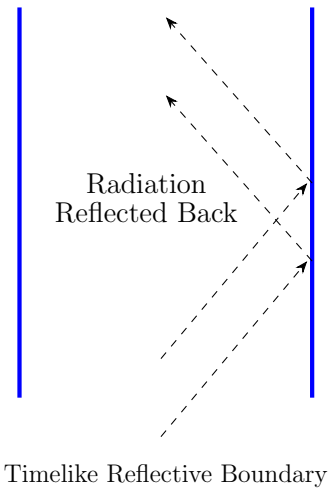

However, the boundary of flat space is fundamentally different. It is \textbf{null infinity} ($\mathscr{I}^+$ for future and $\mathscr{I}^-$ for past, often pronounced \textit{scri-plus/scri-minus}). Unlike the \textit{reflective}, \textit{timelike} wall of AdS, $\mathscr{I}^+$ is a \textit{leaky}, \textit{null hypersurface}(see Fig.\ref{fig:ry}). Radiation, light, gravitational waves or anything massless, crosses it and escapes forever \cite{Madler2016}. Surprisingly, this \textit{leakiness} isn't a bug, in fact it's the defining feature of our reality and the source of the fascinating, infinite-dimensional symmetries that govern it.

\section{Setting the Stage: Bondi-Sachs, $\mathscr{I}^\pm$, and BMS}

If celestial holography is a play, we must now meet its cast of characters in detail. The peculiar geometry of our stage (null infinity), the mathematical language used to describe it (Bondi-Sachs), and the infinite troupe of symmetries (BMS) that dictate the action. This isn't just formalism for formalism's sake; it's about finding the right coordinates to see the hologram.

Think of it this way, describing the expansion of the universe is clumsy in Cartesian coordinates, you need Friedman-Lemaître-Robertson-Walker coordinates. Similarly, describing radiation escaping to infinity is clumsy in standard Minkowski coordinates, one needs coordinates that \textit{follow the outgoing light rays}. This is precisely what the Bondi-Sachs formalism provides, a coordinate system tailored specifically to the \textit{leaky} nature of flat space. It’s the first step in turning a confusing boundary into a manageable holographic hypersurface.

\subsection{The Coordinates of Infinity: Charting $\mathscr{I}^+$ and $\mathscr{I}^-$}

To write equations on the boundary, we need a coordinate chart. The coordinates on null infinity are not arbitrary; they are chosen to reflect the physical reality of light rays traveling to infinity. Their definition differs slightly for the future and past boundaries, mirroring the direction of time's arrow for radiation. To perform holography on a sphere, we use the stereographic coordinates \((z, \bar{z})\) as the standard choice for the celestial sphere because they make the action of Lorentz symmetry manifest.

Consider a unit 2-sphere embedded in \(\mathbb{R}^3\). The standard spherical coordinates are \((\theta, \phi)\). The \textbf{stereographic projection} from the south pole onto the complex plane is given by,
\begin{equation}
z = \cot\left(\frac{\theta}{2}\right) e^{i\phi}, \quad \bar{z} = \cot\left(\frac{\theta}{2}\right) e^{-i\phi}.
\label{eq:stereographic}
\end{equation}
The north pole (\(\theta = 0\)) maps to \(z = \infty\), the south pole (\(\theta = \pi\)) to \(z = 0\), and the equator (\(\theta = \pi/2\)) to the unit circle \(|z| = 1\). The round metric on the sphere is conformally flat in these coordinates,
\begin{equation}
d\Omega^2_{S^2} = d\theta^2 + \sin^2\theta\, d\phi^2 = \gamma_{z\bar{z}}\, dz\, d\bar{z}, \quad \text{where} \quad \gamma_{z\bar{z}} = \frac{2}{(1 + z\bar{z})^2}.
\label{eq:sphere_metric}
\end{equation}
This conformal flatness is super essential, it means that angle-preserving transformations (conformal transformations) on the complex \((z, \bar{z})\)-plane correspond directly to transformations of the sphere.

\subsubsection{Future Null Infinity ($\mathscr{I}^+$): Where Outgoing Light Ends}
For studying processes where radiation \textit{leaves} a system (like a star shining or a black hole evaporating), we use \textbf{retarded coordinates}. The key coordinate is the \textbf{retarded time $u$}. It is defined such that it is constant along outgoing null geodesics (light rays). In flat Minkowski space, $u = t - r$, where $t$ is the usual time and $r$ is a radial distance. On $\mathscr{I}^+$, we describe a point by specifying,
\begin{itemize}
    \item \textbf{$u$}: The retarded time. Physically, it answers, \textit{When did this light ray arrive?}
    \item \textbf{$(z, \bar{z})$}: Stereographic coordinates on the celestial sphere $S^2$. They answer, \textit{From which direction in the sky did this light ray come?}
\end{itemize}
Thus, $\mathscr{I}^+$ is parameterized as $\mathscr{I}^+ \cong \mathbb{R}_u \times S^2$, a cylinder. A light ray with 4-momentum $p^\mu$ that arrives at $\mathscr{I}^+$ is characterized by its energy $\omega$ and its direction $(z, \bar{z})$, where $p^\mu = \omega \, q^\mu(z,\bar{z})$ and $q^\mu$ points to a null direction on the sphere \cite{Pasterski2021}.

\subsubsection{Past Null Infinity ($\mathscr{I}^-$): Where Incoming Light Begins}
The past boundary $\mathscr{I}^-$ is where experiments \textit{begin}; it is the source of in-going states for scattering. Its natural coordinate is the \textbf{advanced time $v$}, constant along \textit{incoming} null geodesics. In flat space, $v = t + r$. On $\mathscr{I}^-$, a point is specified by,
\begin{itemize}
    \item \textbf{$v$}: The advanced time. It answers, \textit{When was this light ray emitted (from the perspective of infinity)?}
    \item \textbf{$(z, \bar{z})$}: The same stereographic coordinates on the celestial sphere. They now answer, \textit{Towards which direction in the sky was this light ray sent?}
\end{itemize}
Therefore, $\mathscr{I}^- \cong \mathbb{R}_v \times S^2$. A crucial detail for scattering is the \textit{antipodal matching} condition \cite{Strominger2017}. To ensure a smooth matching of fields through space-like infinity ($i^0$), a light ray that leaves $\mathscr{I}^-$ at $(v, z, \bar{z})$ is identified with the light ray that arrives at $\mathscr{I}^+$ at $(u, z', \bar{z}')$, where the angles are antipodally related ($z' = -1/\bar{z}$). This matching is essential for defining BMS symmetries consistently across the entire spacetime.

\subsubsection{The Complete Picture: A Scattering Sandwich}
A complete scattering process is thus framed as data on these two boundaries,
\[
\text{In-state on } \mathscr{I}^- \quad \xrightarrow{\text{Non-linear bulk dynamics}} \quad \text{Out-state on } \mathscr{I}^+.
\]
The S-matrix is the unitary operator mapping states defined on $\mathscr{I}^-$ to states defined on $\mathscr{I}^+$. Celestial holography aims at reinterpreting this 4D scattering map as a 2D correlation function on the common celestial sphere $S^2$ that forms the base of both asymptotic cylinders.

\subsection{Lorentz Transformations as Möbius Transformations}
\label{subsec:lorentz-mobius}

The full Lorentz group corresponds to the group of global conformal transformations of the sphere, which are all M\"obius transformations, elements of the $SL(2,\mathbb{C})$ group. A Lorentz transformation in 4D Minkowski spacetime acts on the celestial sphere coordinates $(z, \bar{z})$ as a M\"obius transformation,
\begin{equation}
z \rightarrow z' = \frac{az + b}{cz + d}, \quad
\bar{z} \rightarrow \bar{z}' = \frac{\bar{a}\bar{z} + \bar{b}}{\bar{c}\bar{z} + \bar{d}}
\end{equation}
where the matrix $\begin{pmatrix} a & b \\ c & d \end{pmatrix} \in SL(2,\mathbb{C})$, satisfying $ad - bc = 1$.
\begin{enumerate}
    \item \textbf{Rotations SO(3) subgroup}: 
    These correspond to \textit{unitary M\"obius transformations matrices} with $a = \alpha, b = \beta, c = -\bar{\beta}, d = \bar{\alpha}$, where $|\alpha|^2 + |\beta|^2 = 1$.
    
    \begin{itemize}
        \item[-] \textit{Rotation about the $z$-axis by angle $\phi$}: 
        $a = e^{i\phi/2}, b = 0, c = 0, d = e^{-i\phi/2}$. \\
        This gives: $z \rightarrow e^{i\phi}z$, a simple rotation on the sphere.
        
        \item[-] \textit{Rotation swapping poles (180° rotation about $x$-axis)}: 
        $a = 0, b = 1, c = -1, d = 0$.
        This gives: $z \rightarrow -1/z$.
    \end{itemize}
    
    \item \textbf{Boosts}: These correspond to hyperbolic Möbius transformation matrices.
    \begin{itemize}
        \item[-] \textit{Boost along the $z$-axis}: 
        Parameters: $a = \lambda, b = 0, c = 0, d = 1/\lambda$ for real $\lambda > 0$. \\
        Transformation: $z \rightarrow \lambda z$, a pure dilatation.
        
        \item[-] \textit{Boost along the $x$-axis}: 
        Parameters: $a = \cosh(\frac{\zeta}{2}), \,b = \sinh(\frac{\zeta}{2}),\, c = \sinh(\frac{\zeta}{2}), \,d = \cosh(\frac{\zeta}{2})$, 
        where $\zeta$ is the rapidity. 
        Transformation: $z \rightarrow \dfrac{\cosh(\frac{\zeta}{2})z + \sinh(\frac{\zeta}{2})}{\sinh(\frac{\zeta}{2})z + \cosh(\frac{\zeta}{2})}$.
    \end{itemize}
    
    \item \textbf{Null rotations or special Lorentz transformations}:
    These are transformations that leave a null direction invariant and correspond to parabolic M\"obius transformations matrices.
    
    Parameters: $a = 1, b = \alpha, c = 0, d = 1$ where $\alpha \in \mathbb{C}$. \\
    Transformation: $z \rightarrow z + \alpha$, a simple translation on the celestial sphere. These as will see, are particularly important in the context of BMS supertranslations as will see later.
    
    \item \textbf{The full set}: Any combination of rotations and boosts corresponds to some $SL(2,\mathbb{C})$ matrix, and thus to some M\"obius transformation. The identity is $a = 1,\, b = 0,\, c = 0,\, d = 1$.
\end{enumerate}

This is why the celestial sphere formalism is so powerful as the action of the entire Lorentz group becomes manifest as conformal transformations on the 2D sphere.

\subsection{The Bondi-Sachs Formalism}
The mathematical language tailored for describing radiation leaving an isolated system in general relativity is the Bondi-Sachs formalism \cite{Madler2016}. It employs coordinates $(u, r, y^A)$ adapted to outgoing null geodesics.
\begin{itemize}
    \item $u$ is the \textbf{retarded time}, constant on outgoing null hypersurfaces.
    \item $r$ is a luminosity distance.
    \item $y^A$ ($A=2,3$) are coordinates on the celestial sphere (e.g., $(\theta, \phi)$).
\end{itemize}
In these coordinates, the metric for an asymptotically flat spacetime takes the form,
\begin{equation}
ds^2 = V(u,r,y^A) du^2 + 2du\,dr + r^2 h_{AB}(u,y^A)\left(dy^A - U^A du\right)\left(dy^B - U^B du\right),
% ds^2 = -\frac{V}{r}e^{2\beta} du^2 - 2e^{2\beta}du\,dr + r^2 h_{AB}\left(dy^A - U^A du\right)\left(dy^B - U^B du\right),
\end{equation}
where $V, U^A$ and $h_{AB}$ are functions determined by Einstein's equations. The crucial object is the conformal 2-metric $h_{AB}$ on the sphere. Its traceless part, $C_{AB}(u, y^A)$, is the \textbf{asymptotic shear} tensor which encodes the displacement on the celestial sphere caused by an escaping gravitational wave. Its $u$-derivative, $N_{AB} = \partial_u C_{AB}$, is the \textbf{Bondi news tensor}, which directly encodes the flux of outgoing gravitational radiation \cite{Madler2016}. Future null infinity $\mathscr{I}^+$ is defined as the limit $r \to \infty$ with $u$ and $y^A$ fixed. Topologically, it is a cylinder, $\mathscr{I}^+ \cong \mathbb{R}_u \times S^2$, where ach slice of constant $u$ is a 2-sphere, a snapshot of the sky, and this sphere is our candidate holographic screen.\\

However, to truly appreciate the dynamics at null infinity, we must write the Bondi-Sachs metric in its standard asymptotic form, which explicitly features the key physical quantities \cite{Madler2016, Strominger2017}. In coordinates $(u, r, z, \bar{z})$, where $u$ is retarded time and $(z, \bar{z})$ are complex coordinates on the celestial sphere, the metric reads,

\begin{equation}
\begin{aligned}
ds^2 = & - \left[ 1 - \frac{2M_B(u, z, \bar{z})}{r} + \frac{1}{r^2} \left( \frac{D^z N_{zz} D^{\bar{z}} N_{\bar{z}\bar{z}}}{16} + 4\pi T_{uu} \right) + \dots \right] du^2 \\
& - 2 \left[ 1 + \frac{1}{r^2} \left( \frac{D_z C^{zz} D_{\bar{z}} C^{\bar{z}\bar{z}}}{32} \right) + \dots \right] du \, dr \\
& + \left[ r^2 \gamma_{z\bar{z}} + r C_{zz}(u, z, \bar{z}) + \dots \right] dz^2 \\
& + 2 \left[ r^2 \gamma_{z\bar{z}} - r C_{zz}(u, z, \bar{z}) + \dots \right] d\bar{z} dz \\
& + \left[ \frac{1}{r} \left( \frac{2}{3} N_{z}(u, z, \bar{z}) - \frac{1}{4} \partial_z (C_{zz}C^{zz}) - u \partial_z M_B \right) + \dots \right] du \, dz + \text{c.c.}
\end{aligned}
\label{eq:full_bondi_metric}
\end{equation}

Here, $\gamma_{z\bar{z}} = 2/(1+z\bar{z})^2$ is the metric of the round unit sphere, and $D_z$ is the covariant derivative with respect to it. This formidable expression is worth unpacking, as each term is an important character in our holographic play,

\begin{itemize}
    \item \textbf{Bondi Mass Aspect, \(M_B(u, z, \bar{z})\)}: This is not a single number but a \textit{function on the sphere}. It represents the local energy density at a given angle and retarded time. Its integral over the sphere gives the total Bondi mass, which famously \textit{decreases} as radiation escapes: $\partial_u M \leq 0$.
    \item \textbf{Asymptotic Shear, \(C_{zz}(u, z, \bar{z})\)}: A symmetric, traceless tensor on the sphere, this is the star of the show. It encodes the transverse, radiative degrees of freedom of the gravitational field, that is, the precise imprint of a passing gravitational wave on the celestial sphere. It is related to the supertranslation symmetry.
    \item \textbf{Bondi News, \(N_{zz}(u, z, \bar{z}) = \partial_u C_{zz}\)}: The retarded time derivative of the shear. This is the direct measure of \textit{escaping gravitational radiation flux}. The rate of mass loss is proportional to the square of the news: $\partial_u M_B \propto -\int_{S^2} |N_{zz}|^2$. The news is zero for stationary spacetimes (like Schwarzschild) and non-zero during dynamical events like black hole mergers.
    \item \textbf{Angular Momentum Aspect, \(N_z(u, z, \bar{z})\)}: This function encodes the flux of angular momentum carried away by radiation. It is related to the subleading soft graviton theorem and superrotation symmetry.
\end{itemize}
A similar metric can written at past null infinity \(\mathscr{I}^-\) this time using the \textit{advanced time coordinate} \(v=t+r\) instead of $u$.

The power of this metric lies in its \textit{asymptotic expansion}. The coefficients of different powers of \(1/r\) are not arbitrary; they are constrained by Einstein's equations, which generate an infinite hierarchy of equations relating these physical quantities. The takeaway is profound: the entire future of the spacetime is determined by specifying just two functions on \(\mathscr{I}^+\): the shear \(C_{zz}\) and the Bondi mass aspect \(M_B\) on an initial cut. All other quantities, including the angular momentum aspect, are determined by them via constraints. This is the clearest hint of holography as the two radiative degrees of freedom of $4D$ gravity are packaged into a two-dimensional field \(C_{zz}(u, z, \bar{z})\) living on the null boundary. The Mellin transform we will meet later is essentially a tool to Fourier transform this \(u\)-dependence into a more convenient basis for a conformal theory. One last comment before delving into symmetries, one may ask why such a complicated metric? could we just impose the old good Minkowski fixed metric near infinity? you see, the Bondi-Sachs metric allows for a gravitating object in the bulk of our asymptotically flat universe to radiate gravitational waves. As this radiation will dynamically modify the metric by stretching and displacing as it makes its way towards the null boundary, one should make room for a these changes, that is, the metric should be time-dependent and has more degrees of freedom than our beloved Minkowskian metric. Imposing a rigid metric would prevent radiation and be an artificial reflective boundary with no physical data.

\subsection{The Infinite-Dimensional Playwrights: BMS Transformations}

Having set the stage with the Bondi-Sachs metric, we now meet the directors who can reshape it without changing the physics: the Bondi-Metzner-Sachs (BMS) group. It is not a single symmetry but an infinite tower of them, arising as the \textit{residual diffeomorphisms} that preserve the asymptotic form of the metric in Eq.\eqref{eq:full_bondi_metric} \cite{Madler2016, Strominger2017}. Their action demystifies the celestial sphere and gives physical meaning to the gravitational data recorded on it.

\subsubsection{Supertranslations: The Celestial Time-Keepers}
The most intuitive members are the \textbf{supertranslations}. While an ordinary translation shifts all of retarded time by a constant, \(u \to u + c\), a supertranslation allows this shift to depend on where you are looking in the sky,
\begin{equation}
u \longrightarrow u' = u + \alpha(z, \bar{z}).
\label{eq:supertranslation}
\end{equation}
Here, \(\alpha(z, \bar{z})\) is an \textit{arbitrary smooth function} on the celestial sphere. Now if you decompose the function \(\alpha\) to its spherical harmonics, then the \(l=0\) spherical harmonic (constant \(\alpha\)) is just a global time translation. The \(l=1\) harmonics correspond to ordinary spatial translations known from the Poincaré group. The true novelty lies in the infinitely many modes for \(l \geq 2\).

\textbf{What does it act on?} Supertranslations directly shuffle the fundamental holographic data. Recall the \textit{good cut} we introduced earlier \(u = f_p(z,\bar{z})\) from a bulk point \(p\). A supertranslation maps this to a new cut, say, \(u' = f_p(z,\bar{z}) + \alpha(z,\bar{z})\). It therefore reassigns the \textit{arrival time} of light signals from the same bulk event, depending on their direction. If \(\alpha\) is positive in the northern celestial hemisphere and negative in the south, signals from the north are recorded as arriving later, while those from the south arrive earlier. This is not a mere coordinate relabeling, it is changing the physical phase space. Its action on the shear tensor is given by,
\begin{equation}
\delta_{\alpha} C_{zz}(u, z, \bar{z}) = \alpha(z, \bar{z}) \, \partial_u C_{zz} \textcolor{blue}{- 2 D_z^2 \alpha(z, \bar{z})},
\label{eq:supertrans_on_shear}
\end{equation}
where \(D_z\) is the covariant derivative on the sphere. The first term is just a translation along \(u\). The second term, \(-2 D_z^2 \alpha\), is crucial; it says a supertranslation can \textit{add or subtract} a piece of shear that is constant in time (\(u\)-independent). This time-independent shear corresponds to a \textbf{soft graviton}, that is, a zero-energy, long-wavelength gravitational excitation. As we will see later, the supertranslations are related to \textbf{Weinberg's leading soft graviton theorem}. Physically, they link spacetimes with different configurations of soft graviton emitted from their bulks.

\subsubsection{Superrotations: Stretching the Celestial Sky}
If supertranslations fiddle with celestial clocks (arrival times), \textbf{superrotations} fiddle with the celestial map itself. They extend the Lorentz group which acts on the sphere as global Möbius transformations,

\begin{equation}
   \displaystyle z \to \frac{az+b}{cz+d}, \;\text{where}\quad ad-bc=1,
\end{equation}
to include \textit{local} conformal transformations, that is we allow for Lorentz boosts to be different depending on where you look in the night sky,
\begin{equation}
z \longrightarrow z' = z + \epsilon Y^z(z).
\label{eq:superrotation}
\end{equation}
Here, \(Y^z(z)\) is a \textit{meromorphic vector field}, that is, a 2D-vector field that can have poles on the celestial sphere, generating an infinite set of transformations beyond the well-known six Lorentz transformations. Their generators form a \textit{Virasoro algebra}. A simple example is \(Y^z(z) = z^2\), which is not globally well-defined on the sphere but just to gives a flavor of their action. It stretches angles near the north pole (\(z=0\)) much less than near the equator.

\textbf{What does it act on?} Superrotations directly act on the celestial coordinates \((z, \bar{z})\), meaning they warp the grid of the sky itself. Consequently, they transform the shear tensor as a 2D conformal tensor of weight appropriate to its spin,
\begin{equation}
\delta_{Y} C_{zz}(u, z, \bar{z}) = \mathcal{L}_Y C_{zz} - \frac{1}{2} (D_z Y^z) C_{zz},
\end{equation}
where \(\mathcal{L}_Y\) is the Lie derivative along \(Y\). They also generate a new, \(u\)-dependent component in the transformation of the retarded time, mixing the angular and temporal directions in a non-trivial way. Physically, superrotations are linked to the \textbf{subleading soft graviton theorem} and are associated with changes in the angular momentum aspect \(N_z\) of the spacetime.

\subsubsection{The BMS Group as a Physical Symmetry}
The crucial point is that these are not gauge redundancies but \textit{physical symmetries} with conserved charges. Their Ward identities, \(\langle \text{out} | Q^+ - Q^- | \text{in} \rangle = 0\), are equivalent to the \textbf{soft graviton theorems} \cite{Strominger2017}. The passage of a gravitational wave burst (Bondi news \(N_{zz} \neq 0\)) between early and late times performs a net supertranslation on the spacetime, permanently shifting the shear by \(\Delta C_{zz} = \int N_{zz} du\). This shift, via Eq.\eqref{eq:supertrans_on_shear}, is precisely the \textbf{gravitational memory effect}, that is, a permanent displacement of inertial detectors that can, in principle, be measured by gravitational wave observatories like LIGO/Virgo.

This infinite-dimensional symmetry is the holographic gold. Its associated Ward identities are mathematically equivalent to the famous \textbf{soft theorems} (e.g., Weinberg's soft graviton theorem). This profound link means that soft particles (zero-energy photons/gravitons) are the Goldstone bosons of spontaneously broken asymptotic symmetries \cite{Strominger2017}. The \textit{leakiness} of the boundary and the inevitable \textit{leaky emission} of soft quanta in any scattering process are two sides of the same coin.

To sum up, the BMS group is the infinite-dimensional symmetry of the leaky boundary. \textit{Supertranslations}, \(\alpha(z,\bar{z})\), reshuffle the arrival times of signals on the celestial sphere, while \textit{Superrotations}, \(Y^z(z)\),  stretch and warp the celestial map itself. They act directly on the radiative data, the shear \(C_{zz}\), and through it, on the very structure of the holographic screen. Their existence forces the radical conclusion that the Hilbert space of quantum gravity in flat space is organized not under the tiny Poincaré group, but under this vast, celestial symmetry.

\subsection{Good Cuts, Light Cones, and Holographic Data}

\begin{figure}[h!]
    \centering
    \begin{tikzpicture}[scale=1.2]
        \draw[thick, color=blue] (0,4) ellipse (2cm and 0.6cm); 
        \draw[thick, color=blue] (-2,0) -- (-2,4); 
        \draw[thick, color=blue] (2,0) -- (2,4);   
        \node[blue, above] at (0,3.7) {\(\mathscr{I}^+\)};
        \node[blue, right] at (2,2) {\(\mathbb{R}_u \times S^2\)};

        \filldraw[black] (0,1.5) circle (2pt) node[left] {\(p\)};
        \draw[blue, thick, dashed] (0,1.5) -- (-1.5, 0.1);
        \draw[thick, dashed] (0,1.5) -- (1.5, 0.1);
        \draw[thick, dashed] (0,1.5) -- (-1.5, 3.9);
        \draw[thick, dashed] (0,1.5) -- (1.5, 3.9);

        \draw[ultra thick, dashed, blue] (2,0.) arc (0:180:2cm and 0.6cm);
        \draw[ultra thick, very thick,blue] (-2,0.) arc (180:360:2cm and 0.6cm); 
        
         \draw[thick] (0,0) ellipse (1.5cm and 0.4cm); 
       
        \draw[thick, red ] (0,4) ellipse (1.5cm and 0.4cm); 
        \node[align=center,black] at (0, 4.9) {Good Cut \(C^+(p)\): \(u = f_p(z, \bar{z})\)};
        \draw[->>,black,thick](1.7,4.7) arc[
        start angle=70,
        end angle=-90,
        x radius=0.7cm,
        y radius =0.3cm
        ];
        
    \end{tikzpicture}
    \caption{\small \textbf{Good Cuts from Light Cones.} A bulk event \(p\) generates a future light cone (black, dashed). Its intersection with null infinity \(\mathscr{I}^+\) is a two-dimensional surface called a good cut \(C^+(p)\), which can be described by an arrival time function \(f_p\) on the celestial sphere. The complete set of such cuts encodes the bulk's causal structure.}
    \label{fig:good_cuts}
\end{figure}
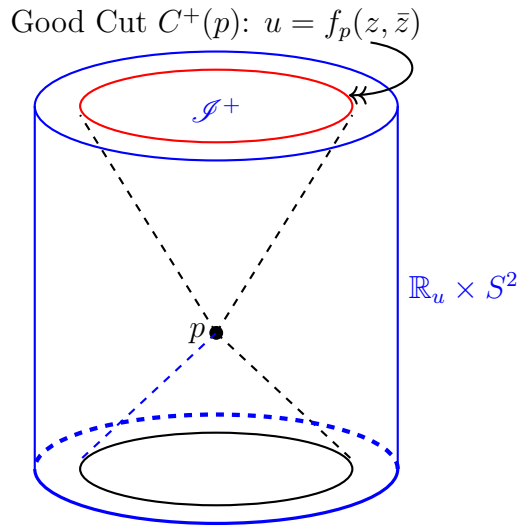

The celestial sphere is more than a screen for projections. In fact, it is an archival surface where the causal history of the bulk is recorded. This archival process is captured by the concept of \textit{good cuts}. Consider a point \(p\) in the bulk of an asymptotically flat spacetime. Its future light cone, that is, the set of all null rays emanating from \(p\), expands outwards and eventually intersects future null infinity \(\mathscr{I}^+\). This intersection is not a single point but a two-dimensional cross-section, called a \textit{cut}, of the cylindrical \(\mathscr{I}^+\) \cite{Esposito2024}. This specific cut \(C^+(p)\) is called a \textit{good cut} (or light cone cut) associated with the bulk event \(p\) (see Fig.\ref{fig:good_cuts}).

The geometry is quite elegant since the light cone from \(p\) defines, for each celestial direction \((z, \bar{z})\), the precise retarded time \(u = f_p(z, \bar{z})\) at which the light ray arrives at \(\mathscr{I}^+\). Therefore, a good cut is described by a function on the sphere such as,
\[
C^+(p): \quad u = f_p(z, \bar{z}).
\]
The entire set of good cuts \(\{C^+(p) \text{ for all bulk events } p\}\) constitutes the \textit{holographic data}. It encodes the bulk's causal structure, meaning which events in the bulk can be connected by light rays, just by analyzing patterns in arrival times on the boundary. This the very essence of holography!

Now, the drama unfolds. The spacetime is dynamical. When a gravitational wave represented by a burst of Bondi news \(N_{zz}\) passes through the bulk, it \textit{warps} the geometry. This warping changes the trajectories of subsequent null rays. Consequently, the pattern of good cuts on \(\mathscr{I}^+\) is \textit{shuffled}. A specific good cut \(C^+(p)\) associated with an event after the wave will be displaced relative to where it would have been had the wave not passed. This displacement is precisely encoded in a \textbf{BMS supertranslation}.

A supertranslation, as we've seen, acts as \(u \to u + \alpha(z,\bar{z})\). If the passage of radiation induces a net change in the asymptotic shear by an amount \(\Delta C_{zz} = \int N_{zz} du\), then the corresponding shuffling of all good cuts is described by a supertranslation with function \(\alpha(z,\bar{z})\) satisfying \(D_z^2 \alpha \propto \Delta C_{zz}\). In this way, the \textit{leaky} flux of news from the bulk (\(N_{zz} \neq 0\)) directly manifests as a \textit{reorganization} of the holographic data i.e. the good cuts, via the infinite-dimensional BMS symmetry \cite{Esposito2024}. The conservation of BMS charge ensures this shuffling is consistent and measurable and gives rise to the gravitational memory effect.

The profound claim at the heart of celestial holography is that this data is not only sufficient but also \textit{necessary}. In other words, the set of all good cuts on \(\mathscr{I}^+\) is \textit{believed to be sufficient} to reconstruct the \textbf{conformal structure} of the entire bulk spacetime. In other words, knowing which null geodesics intersect where and when allows one to reconstruct the network of causal relationships, which in turn determines the spacetime metric up to an overall conformal factor. This establishes the direct holographic correspondence such that the conformal geometry of the four-dimensional bulk is encoded in the arrangement of two-dimensional surfaces (cuts) on its three-dimensional null boundary.

\subsection{Some Twisted Taste: Light Rays as Primary}

The philosophy that spacetime events should be derived from more primitive structures finds its ultimate expression in \textbf{twistor theory}, pioneered by Roger Penrose \cite{Penrose1967, WikipediaTwistor}. Remarkably, the celestial holography program resonates deeply with this twistor philosophy, often viewing it as a concrete realization of twistor ideas in a holographic context.

The core concept of twistor theory is a radical swap of primitives. In standard physics, a spacetime point is fundamental, and a light ray, a \textit{null geodesic}, is a set of points. In twistor theory, a \textit{twistor} is an object that represents a \textbf{light ray} (or any spinning massless particle) as a fundamental entity \cite{WikipediaTwistor}. Minkowski spacetime points then become \textit{derived} secondary objects. They correspond to the set of all light rays passing through that point, which in twistor space is a complex projective line (\(\mathbb{CP}^1\)). This is encapsulated in the \textit{incidence relation},
\[
\omega^A = i x^{AA'} \pi_{A'}
\]
where \((\omega^A, \pi_{A'})\equiv Z^{\alpha}\) is a twistor and \(x^{AA'}\) is a spacetime point \cite{WikipediaTwistor}.

This reversal of logic is perfectly aligned with the geometric picture of celestial holography. Our holographic data represented by all the good cuts are precisely the sets of light rays labeled by their celestial direction \((z,\bar{z})\) and arrival time \(u\) that define events on \(\mathscr{I}^+\). In fact, a point on \(\mathscr{I}^+\) is specified by a pair \((u, z, \bar{z})\), which is equivalent to specifying a unique \textit{asymptotic light ray}. The celestial sphere itself is the space of asymptotic light ray directions, which in twistor terms is the projective spinor space parametrized by \(\pi_{A'}\).

The connection becomes a powerful computational tool via the \textbf{Penrose transform}. This transform establishes that solutions to massless field equations, for example, the linearized Einstein equations for gravitational waves in spacetime correspond to holomorphic functions (more precisely, cohomology classes) on regions of twistor space \cite{Penrose1967, WikipediaTwistor, TwistorLi}. In celestial holography, the Mellin-transformed scattering amplitudes, the celestial amplitudes \(\mathcal{C}_n\), have been shown to have natural and often simpler descriptions in twistor space \cite{Mason2011}. The conformal basis we use for celestial operators is intimately related to the helicity basis provided by twistors.

Thus, celestial holography can also be viewed as a bridge as it takes the foundational twistor idea that light rays are primary and implements it within a \textit{holographic framework}. The CCFT is the theory that organizes the data of these asymptotic light rays. \textit{While twistor theory seeks to build spacetime from twistor space, celestial holography seeks to describe quantum gravity in spacetime via a CFT on the space of asymptotic twistors, the celestial sphere}. This synergy makes twistor theory a rich source of mathematical techniques and conceptual insights for the celestial program.

\section{The Holographic Camera: The Mellin Transform and Celestial Amplitudes}

We have a stage ($\mathscr{I}^+$) and its symmetries (BMS). The central question of celestial holography is \textit{How do we explicitly map a 4D scattering event to a 2D correlator?} The answer involves a clever change of basis, powered by the \textbf{Mellin transform}.

\subsection{From Momentum Eigenstates to Conformal Primaries}
In standard quantum field theory, asymptotic particles in scattering amplitudes are described by plane waves, which are eigenstates of \textit{momentum}. For a massless particle, we can write the null 4-momentum as,
\begin{equation}
p^\mu = \omega \, q^\mu(z, \bar{z}), \quad \text{where } q^\mu = \frac{1}{\sqrt{2}}\left(1 + z\bar{z}, z + \bar{z}, -i(z-\bar{z}), 1 - z\bar{z}\right).
\end{equation}
Here, $\omega > 0$ is the energy, and $(z, \bar{z})$ are the stereographic coordinates on the celestial sphere ($S^2 \cong$ complex plane).

For holography, we want operators that transform as primaries under the action of Lorentz transformations (which act as Möbius transformations on $(z,\bar{z})$). We therefore need eigenstates of \textbf{boosts}, not of translations like the plane waves. This is achieved by trading the energy $\omega$ for a \textbf{conformal dimension} $\Delta$ via a Mellin transform of creation/annihilation operators in the bulk,
\begin{equation}
\mathcal{O}_{\Delta, J}(z, \bar{z}) = \int_{0}^{\infty} \frac{d\omega}{\omega} \, \omega^{\Delta} \, a(\omega, z, \bar{z}; J).
\label{eq:mellin_transform}
\end{equation}
The operator $\mathcal{O}_{\Delta, J}(z, \bar{z})$ is a \textbf{celestial operator}. The parameter $\Delta$ is its conformal (or boost) weight, and $J$ is its spin. Under a Lorentz transformation mapping $\displaystyle
 z \to \frac{az+b}{cz+d}$, it transforms as a 2D primary field,
\begin{equation}\label{eq:primary_operator_law}
\mathcal{O}_{\Delta, J}\left(z, \bar{z}\right) \to \left| \frac{\partial z'}{\partial z} \right|^{-\Delta} \left( \frac{\partial z'}{\partial z} \bigg/ \frac{\bar{\partial} \bar{z}'}{\bar{\partial} \bar{z}} \right)^{J/2} \mathcal{O}_{\Delta, J}(z', \bar{z}').
\end{equation}

Celestial operators are called \textit{primary} because they transform \textit{homogeneously} under the global conformal transformations. That means when we apply a small coordinate change $(z,\bar{z})\to (z',\bar{z}')$, the operator picks up a factor that is a power of the Jacobian of the transformation as seen in Eq.\eqref{eq:primary_operator_law}.

\subsection{Celestial vs. Momentum Operators: Eigenstates of Different Symmetries}

A common source of confusion when encountering celestial holography is the relationship between the new celestial operators and the familiar creation/annihilation operators of quantum field theory. The distinction is fundamental as they create states that are eigenstates of \textit{different symmetry generators}.

\subsubsection{Momentum Basis: Plane Waves}
In the standard formulation of scattering theory, asymptotic one-particle states are created from the vacuum by momentum-space creation operators,
\[
|\omega, z, \bar{z}; J\rangle \equiv a^\dagger(\omega, z, \bar{z}; J) |0\rangle.
\]
These states \(|\omega, z, \bar{z}; J\rangle\) are \textbf{momentum eigenstates}. They diagonalize the generators of spacetime \textit{translations} (\(P_\mu\)). A plane wave \(e^{i p\cdot x}\) is the wavefunction of such a state.

\subsubsection{Celestial Basis: Conformal Primaries}
The celestial operators, defined via the Mellin transform, create a different class of states,
\[
|\Delta, z, \bar{z}; J\rangle_{\text{cel}} \equiv \mathcal{O}_{\Delta, J}(z, \bar{z}) |0\rangle = \int_{0}^{\infty} d\omega \; \omega^{\Delta-1} \, |\omega, z, \bar{z}; J\rangle.
\]
These states \(|\Delta, z, \bar{z}; J\rangle_{\text{cel}}\) are \textbf{boost eigenstates}. They diagonalize the generators of Lorentz \textit{boosts} (and dilatations). Their wavefunctions are the conformal primary wavefunctions denoted \(\phi_\Delta^{\pm}(X;z,\bar{z})\) not plane waves.

A single celestial state is a \textit{superposition} of plane waves of all energies \(\omega\), weighted by the kernel \(\omega^{\Delta-1}\). A boost, which would scramble a plane wave of fixed energy into a mixture of different energies, simply multiplies a celestial state by a pure number. This is why the celestial basis is natural for holography. Lorentz transformations act geometrically on \((z,\bar{z})\) and multiplicatively on \(\Delta\).

\subsubsection{The Basis Change}
Thus, the Mellin transform is precisely a unitary (or at least orthogonal) \textbf{change of basis} in the asymptotic Hilbert space,
\[
\text{Momentum Eigenstates } |\omega, z, \bar{z}\rangle \quad \xrightleftharpoons[\text{Transform}]{\text{Mellin}} \quad \text{Boost Eigenstates } |\Delta, z, \bar{z}\rangle_{\text{cel}}.
\]
The scattering S-matrix, which is an operator mapping the asymptotic in-Hilbert space to the out-Hilbert space, can be expressed in either basis. Calculating its matrix elements in the boost basis yields the celestial amplitudes,
\[
{}_{\text{cel}}\langle \text{out} | S | \text{in} \rangle_{\text{cel}} = \langle \mathcal{O}_{\Delta_1} \cdots \mathcal{O}_{\Delta_n} \rangle_{\text{CCFT}}.
\]

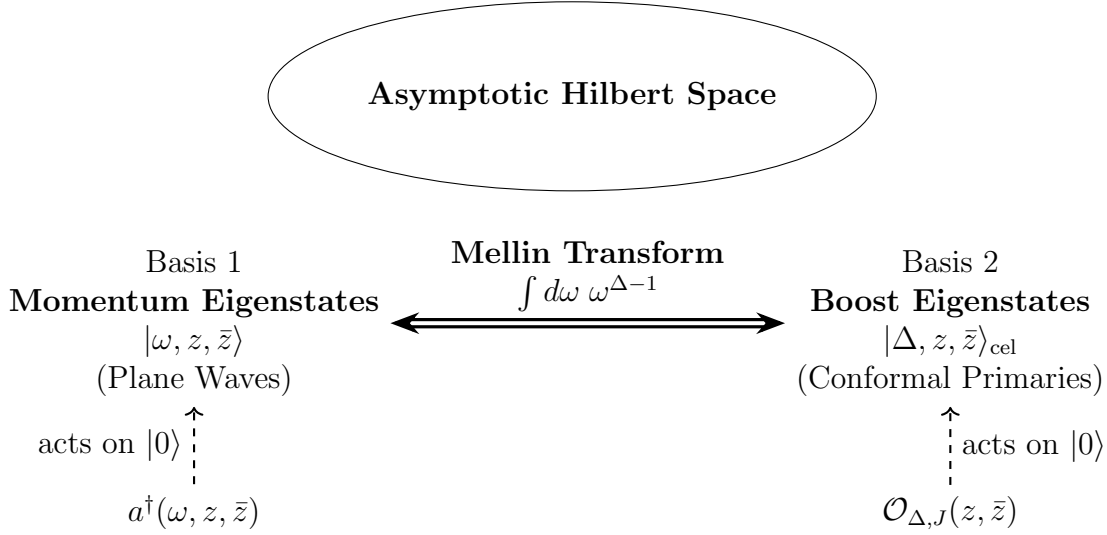
\begin{figure}[h!]
    \centering
    \begin{tikzpicture}[node distance=2cm,scale=0.6]
        \node (hilbert) [draw, ellipse, minimum width=4cm, minimum height=2.5cm] {\textbf{Asymptotic Hilbert Space}};
        
        \node (momentumbasis) [left of=hilbert, xshift=-3cm,yshift=-3cm, align=center] {Basis 1\\ \textbf{Momentum Eigenstates}\\ $|\omega, z, \bar{z}\rangle$\\ (Plane Waves)};
        \node (boostbasis) [right of=hilbert, xshift=3cm,yshift=-3cm, align=center] {Basis 2\\ \textbf{Boost Eigenstates}\\ $|\Delta, z, \bar{z}\rangle_{\text{cel}}$\\ (Conformal Primaries)};
        
        \node (adagger) [below of=momentumbasis, yshift=-0.5cm] {$a^\dagger(\omega, z, \bar{z})$};
        \node (ocal) [below of=boostbasis, yshift=-0.5cm] {$\mathcal{O}_{\Delta, J}(z, \bar{z})$};
        
        \draw[->, thick, dashed] (adagger.north) -- node[left, pos=0.5] {acts on $|0\rangle$} (momentumbasis.south);
        \draw[->, thick, dashed] (ocal.north) -- node[right, pos=0.5] {acts on $|0\rangle$} (boostbasis.south);
        
        \draw[<->, very thick, >=Stealth, double, double distance=1.5pt ](momentumbasis.east) -- node[above, midway, align=center] {\textbf{Mellin Transform}\\ $\small \int d\omega\;\omega^{\Delta-1}$} (boostbasis.west);
        
    \end{tikzpicture}
    \caption{\textbf{Two Bases for the Same Hilbert Space.} The asymptotic Hilbert space of single-particle states can be spanned either by momentum eigenstates (created by $a^\dagger$) or by boost eigenstates (created by $\mathcal{O}_{\Delta}$). The Mellin transform is the explicit change-of-basis integral that relates them. The S-matrix can be evaluated in either basis, but the boost basis makes conformal symmetry manifest.}
    \label{fig:two_bases}
\end{figure}

\subsubsection{Why is the Boost Basis Useful for Holography?}
The holographic principle suggests that the degrees of freedom in a region of spacetime are encoded on its boundary. The boundary of asymptotically flat spacetime, $\mathscr{I}^+$, is naturally endowed with an action of the Lorentz group (and its BMS extension). The boost basis is adapted to this boundary symmetry,
\begin{itemize}
    \item In the momentum basis, a Lorentz boost mixes the energy \(\omega\) and the direction \((z,\bar{z})\) in a complicated, non-diagonal way.
    \item In the celestial (boost) basis, a Lorentz boost acts \textit{diagonally} on the state: it simply multiplies it by a factor and shifts \((z,\bar{z})\) via a Möbius transformation. The quantum number \(\Delta\) is the boost eigenvalue.
\end{itemize}
This is why correlation functions in the boost basis, \textit{celestial amplitudes}, transform like those of a 2D CFT. The operators \(\mathcal{O}_{\Delta,J}(z,\bar{z})\) are crafted to be \textbf{primary operators} under this action. Therefore, celestial operators do not create new kinds of particles, they create a new, symmetry-adapted basis of states for the same particles, perfectly suited to a holographic description on the celestial sphere.

\subsection{Celestial Amplitudes: The Core Dictionary}
\begin{figure}[H]
    \centering
    \begin{tikzpicture}[node distance=3.2cm, thick,scale=0.8]
        \node (bulk) [draw, rectangle, minimum width=3cm, minimum height=2cm, align=center] {\textbf{4D Bulk Physics}\\Scattering Amplitude\\$\mathcal{A}_n(\omega_i, z_i)$};
        \node (mellin) [draw, diamond, aspect=2, minimum width=1.5cm, right of=bulk, xshift=2.5cm, align=center] {\textbf{Mellin}\\Transform\\$\int d\omega \, \omega^{\Delta-1}$};
        \node (cft) [draw, rectangle, minimum width=3cm, minimum height=2cm, right of=mellin, xshift=2.5cm, align=center] {\textbf{2D Boundary CFT}\\Correlator\\$\textcolor{red}{\mathcal{C}_n}\equiv\langle \mathcal{O}_{\Delta_1}(z_1) \cdots \rangle$};

        \draw[->, >=Stealth, very thick] (bulk) --  (mellin);
        \draw[->, >=Stealth, very thick] (mellin) --  (cft);
        \node[below of=bulk, yshift=1.5cm, align=center] {\small \textbf{Traditional S-Matrix}\\(Standard QFT)};
        \node[below of=cft, yshift=1.5cm, align=center] {\small \textbf{Celestial CFT}\\(Conjectured Hologram)};
    \end{tikzpicture}
    \caption{\small \textbf{The Holographic Map.} The Mellin transform serves as the crucial change of basis, converting a standard momentum-space amplitude into a celestial amplitude, which is interpreted as a correlator in a two-dimensional conformal field theory on the celestial sphere.}
    \label{fig:holographic_map}
\end{figure}
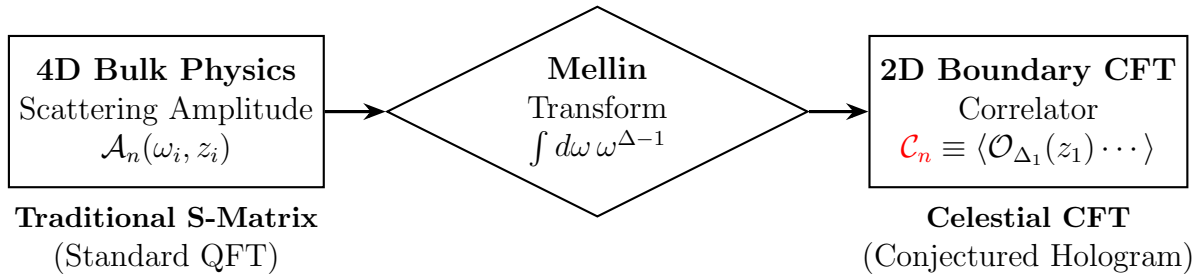

Consider an $n$-particle momentum-space scattering amplitude $\mathcal{A}_n\left(\{\omega_i, z_i, \bar{z}_i; J_i\}\right)$. To obtain its holographic counterpart, the \textbf{celestial amplitude}, we simply Mellin-transform each external leg according to Eq.\eqref{eq:mellin_transform},
\begin{equation}
{\mathcal{C}}_n\left(\{\Delta_i, z_i, \bar{z}_i; J_i\}\right) = \left( \prod_{i=1}^n \int_0^\infty \frac{d\omega_i}{\omega_i} \omega_i^{\Delta_i} \right) \mathcal{A}_n\left(\{\omega_i, z_i, \bar{z}_i; J_i\}\right).
\label{eq:celestial_amplitude}
\end{equation}
This object ${\mathcal{C}}_n$ is then \textit{defined} as the correlation function of celestial operators in the \textit{still} hypothetical Celestial Conformal Field Theory (CCFT),
\begin{equation}
{\mathcal{C}}_n\left(\{\Delta_i, z_i, \bar{z}_i\}\right) \quad \stackrel{\text{\tiny dict}}{\equiv} \quad \langle \mathcal{O}_{\Delta_1, J_1}(z_1, \bar{z}_1) \cdots \mathcal{O}_{\Delta_n, J_n}(z_n, \bar{z}_n) \rangle_{\text{CCFT}} .
\end{equation}
This is the central conjecture of celestial holography. The Mellin transform is the \textit{holographic camera} that repackages 4D scattering data into a 2D conformal correlator on the boundary celestial sphere, Fig\ref{fig:holographic_map}. The soft theorems derived from BMS symmetry now manifest as conformal Ward identities in this CCFT.

\subsection{The Conformal dimension $\Delta$ revisited: Boost Eigenvalue and Scaling Dimension}

In a two-dimensional conformal field theory, a primary operator $\mathcal{O}(z, \bar{z})$ is defined by its transformation law under a conformal/Möbius map $\displaystyle z \to z' = \frac{az+b}{cz+d}$:
\begin{equation}
\mathcal{O}(z, \bar{z}) \to \left( \frac{\partial z'}{\partial z} \right)^{h} \left( \frac{\partial \bar{z}'}{\partial \bar{z}} \right)^{\bar{h}} \mathcal{O}(z'(z), \bar{z}'(\bar{z})).
\label{eq:primary_law}
\end{equation}
The numbers $h$ and $\bar{h}$ are called the conformal weights. Their sum $\Delta = h + \bar{h}$ is the conformal/scaling dimension, which dictates how the operator responds to a uniform dilatation ($z \to \lambda z$). Their difference $J = h - \bar{h}$ is the spin.

In celestial holography, the parameter $\Delta$ that appears in the Mellin transform
\begin{equation}
\mathcal{O}_{\Delta, J}(z, \bar{z}) = \int_0^\infty d\omega\, \, \omega^{\Delta-1} \, a(\omega, z, \bar{z}; J)
\end{equation}

is precisely this scaling dimension. Physically, $\Delta$ is the eigenvalue of the particle state under Lorentz boosts. To see this, consider a Lorentz boost along the $z$-axis. In celestial coordinates, this acts as a dilatation on the complex plane, $z \to \lambda z$, $\bar{z} \to \lambda \bar{z}$, for some real $\lambda$. Under such a boost,
the energy of a massless particle transforms as $\omega \to \lambda^{-1} \omega$. The creation operator $a(\omega, z, \bar{z}; J)$ picks up a phase factor from its spin $J$, but also inherits the scaling of $\omega$. Now, examine the Mellin-transformed celestial operator,
\begin{equation}
\mathcal{O}_{\Delta, J}(\lambda z, \lambda \bar{z}) = \int_0^\infty \frac{d\omega}{\omega} \, \omega^{\Delta} \, a(\omega, \lambda z, \lambda \bar{z}; J).
\end{equation}

By changing the integration variable to $\omega' = \lambda \omega$ and using how $a(\omega, z, \bar{z}; J)$ transforms under a boost, one finds, up to a spin phase,
\begin{equation}  
\mathcal{O}_{\Delta, J}(\lambda z, \lambda \bar{z}) = \lambda^{-\Delta} \mathcal{O}_{\Delta, J}(z, \bar{z}).
\end{equation}

Comparing this to the conformal transformation law \eqref{eq:primary_law} for a pure dilatation ($h=\bar{h}=\Delta/2$), we see they match perfectly. Therefore, the celestial operator $\mathcal{O}_{\Delta, J}(z, \bar{z})$ is a conformal primary because it is constructed as a boost eigenstate. The Mellin transform kernel $\omega^{\Delta-1}$ is the unique choice that performs this change of basis from energy ($\omega$) eigenstates to boost ($\Delta$) eigenstates.

In unitary CFTs, $\Delta$ is real and bounded from below. In celestial holography, the principal series representation, where $\Delta = 1 + i\lambda$ ($\lambda \in \mathbb{R}$), is particularly important. This corresponds to a complete, orthogonal basis for plane waves, analogous to the decomposition of plane waves into spherical waves. Operators with real $\Delta$ correspond to wavefunctions with specific power-law fall-offs in the bulk.

One may think of the conformal weight $\Delta$ as the celestial currency for energy. It is the charge that tells you how an operator scales when you \textit{zoom in} on the celestial sphere, which is exactly what a Lorentz boost does. This elegant correspondence is why the infinite-dimensional BMS symmetries which do include Lorentz boosts as a subgroup act as conformal transformations on the celestial CFT.

\section{The Conserved Leak: Ward Identities, Soft Theorems, and Supertranslation Charges}

At the heart of celestial holography lies a magnificent equivalence that links the \textit{geometry} of the boundary to the \textit{dynamics} of the bulk. This is the correspondence between BMS Ward identities and soft theorems which a direct consequence of the \textbf{leaky} nature of flat spacetime.

\subsubsection{The Leakiness Correspondence}
$\mathscr{I}^+$ is a \textit{leaky} boundary since gravitational radiation, through the Bondi news $N_{zz}$, carries energy, momentum, and information out through it. In the bulk, any scattering process involving accelerated masses is also \textit{leaky} as it must emit a shower of very low-energy \textit{soft} gravitons. The magic is that these are not two different leaks in the contrary they are the \textbf{same leak} described in two different languages. The radiation crossing the boundary turns out to be the soft graviton cloud dressing the asymptotic outgoing/ingoing states.

\subsubsection{The Supertranslation Charge}
The generator of a supertranslation $u \rightarrow u + \alpha(z,\bar{z})$ is a conserved charge $Q[\alpha]$. On a cut of $\mathscr{I}^+$, it is constructed from the asymptotic data \cite{Strominger2017},
\begin{equation}
Q[\alpha] = \frac{1}{4\pi G} \int_{S^2} d^2z \, \gamma_{z\bar{z}} \, \alpha(z,\bar{z}) \, m_B(z,\bar{z}) + \text{(terms from $C_{zz}$)}.
\end{equation}
\textbf{One can ask what carries this charge?} It is not a localized particle. The charge is carried by,
\begin{enumerate}
    \item The \textbf{mass-energy distribution} at infinity, encoded in the Bondi mass aspect $m_B(z,\bar{z})$, hard part.
    \item The configuration of the \textbf{asymptotic gravitational field itself}, specifically the shear $C_{zz}$, soft part.
\end{enumerate}
When gravitational waves pass, they change the shear and therefore carry supertranslation charge from the bulk to infinity.

\subsubsection{The Ward Identity and its Physical Meaning}
The fundamental quantum-mechanical statement is that the total supertranslation charge is conserved between the far past and the far future. For any admissible scattering process, we must have,
\begin{equation}
\langle \text{out} |\, \left( Q^+[\alpha] - Q^-[\alpha] \right)\, | \text{in} \rangle = 0.
\label{eq:ward_identity}
\end{equation}
This Ward identity has a breathtaking consequence. When you carefully compute the action of the charges $Q^\pm$ on the $in/out$ states, Eq.\eqref{eq:ward_identity} \textit{forces} a specific factorization of the amplitude. This factorization is precisely Weinberg's leading soft graviton theorem,
\begin{equation}
\lim_{\omega \to 0} \langle \text{out} + \text{soft graviton}(\omega, z) | \text{in} \rangle = \left[ \frac{\kappa}{2} \sum_{k} \frac{p_k^\mu p_k^\nu \epsilon_{\mu\nu}}{p_k \cdot q} \right] \langle \text{out} | \text{in} \rangle.
\label{eq:soft_theorem}
\end{equation}

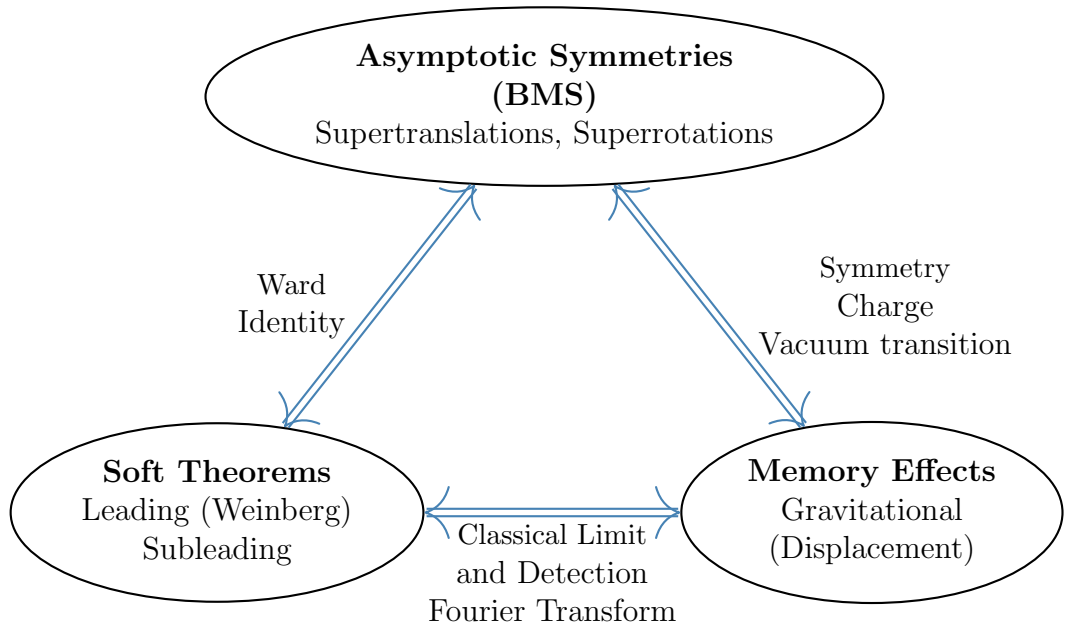
\begin{figure}[!ht]
    \centering
    \begin{tikzpicture}[
        node distance=2cm,
        main/.style={draw, ellipse, minimum width=3.2cm, minimum height=1.8cm, align=center, thick, fill=white},
        arrow/.style={<->,scale=1.5, thick, double, double distance=2pt,scale=0.4}
    ]
        \node[main] (symmetry) at (90:3) {\textbf{Asymptotic Symmetries}\\ \textbf{(BMS)}\\ Supertranslations, Superrotations};
        \node[main] (soft) at (210:5) {\textbf{Soft Theorems}\\ Leading (Weinberg)\\ Subleading};
        \node[main] (memory) at (330:5) {\textbf{Memory Effects}\\ Gravitational\\ (Displacement)};

        \draw[arrow, color=SteelBlue] (symmetry) -- node[left, pos=0.5, align=center, text=black,xshift=-0.3cm] {\small Ward\\Identity} (soft);
        \draw[arrow, color=SteelBlue] (soft) -- node[below, pos=0.5, align=center, text=black] {\small Classical Limit\\and Detection\\ Fourier Transform} (memory);
        \draw[arrow, color=SteelBlue] (memory) -- node[right,xshift=0.5cm, pos=0.5, align=center, text=black] {\small Symmetry\\Charge\\ Vacuum transition} (symmetry);

    \end{tikzpicture}
    \caption{\small \textbf{The Infrared Triangle.} This diagram illustrates the fundamental, three-way equivalence that is central to celestial holography. The infinite-dimensional asymptotic symmetries (BMS) of flat spacetime imply conservation laws (Ward identities). These Ward identities are mathematically equivalent to the universal soft theorems governing the emission of low-energy particles. The same symmetries and soft emission predict observable memory effects manifesting as permanent changes in the relative positions of detectors. These three concepts are not independent as they are three facets of the same infrared structure of quantum gravity.}
    \label{fig:infrared_triangle}
\end{figure}

This implies the following physical connections, Fig.\ref{fig:infrared_triangle},
\begin{itemize}
    \item The soft graviton in Eq.\eqref{eq:soft_theorem} is the \textit{Goldstone boson} of spontaneously broken supertranslation symmetry.
    \item The theorem is a \textit{conservation law} stating that the sum of the soft factors (weighted by $\alpha(z)$) must vanish, ensuring the net flux of supertranslation charge is balanced.
    \item This directly implies the \textit{gravitational memory effect}. The net change in shear between early and late times, $\Delta C_{zz} = \int N_{zz} du$, is a supertranslation, $\Delta C_{zz} = -2 D_z^2 \alpha$. The displacement of test masses measures \textit{the memory of the charge} that has leaked through.
\end{itemize}

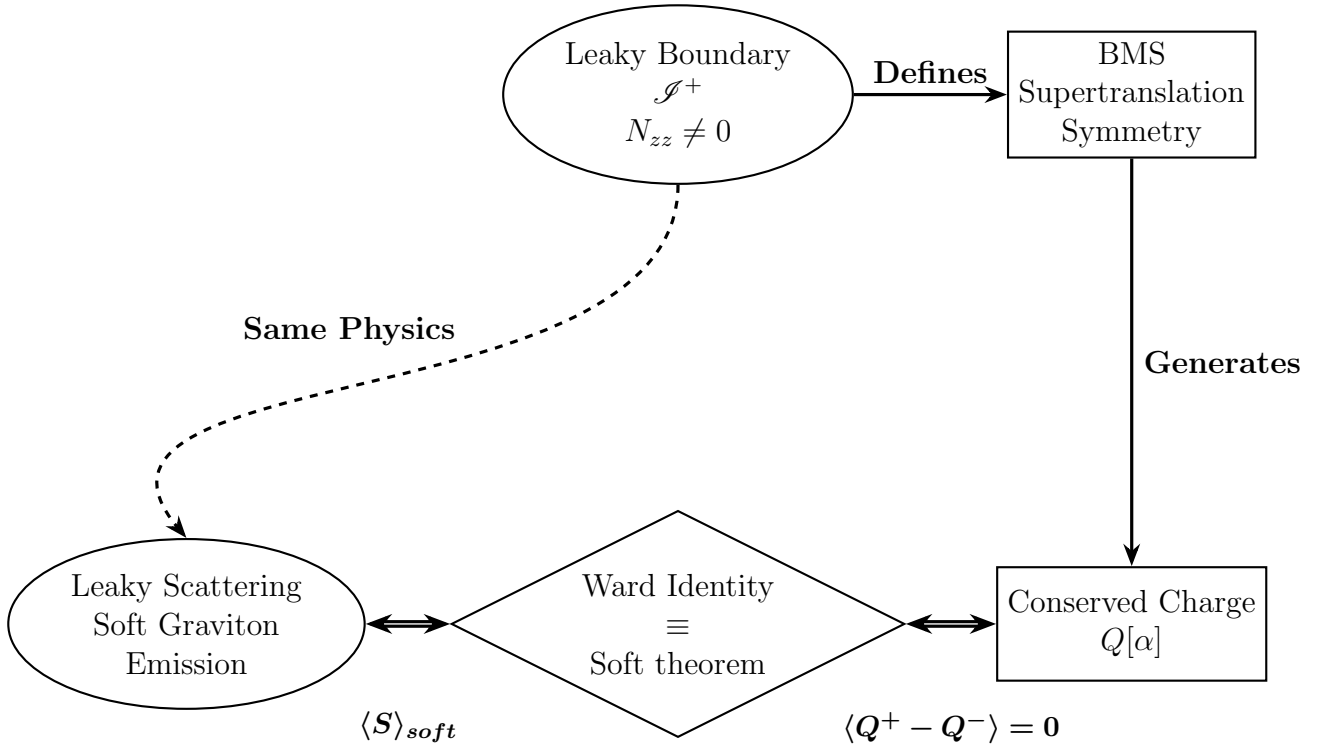
\begin{figure}[h!]
    \centering
    \begin{tikzpicture}[node distance=6cm, thick]
        \node (leaky) [draw, ellipse, minimum width=3cm, minimum height=1.5cm,  align=center] {Leaky Boundary\\$\mathscr{I}^+$\\$N_{zz} \neq 0$};
        \node (bms) [draw, rectangle, right of=leaky, minimum width=3cm, minimum height=1.5cm, align=center] {BMS\\Supertranslation\\Symmetry};
        \node (charge) [draw, rectangle, below of=bms, yshift=-1cm, minimum width=3cm, minimum height=1.5cm,  align=center] {Conserved Charge\\$Q[\alpha]$};
        \node (ward) [draw, diamond, aspect=2, left of=charge, minimum width=2cm,  align=center] {Ward Identity\\$\equiv$ \\Soft theorem};
        \node (soft) [draw, ellipse, left of=ward, xshift=-0.5cm, minimum width=3cm, minimum height=1.5cm,  align=center] {Leaky Scattering\\Soft Graviton\\Emission};

        \draw[->, >=Stealth, very thick] (leaky) -- node[above] {\textbf{Defines}} (bms);
        \draw[->, >=Stealth, very thick] (bms) -- node[right] {\textbf{Generates}} (charge);
        \draw[<->, >=Stealth, very thick, double, double distance=0.5pt] (charge) -- node[below, yshift=-1cm] {\small{$\bm{\langle Q^+ - Q^- \rangle = 0}$}} (ward);
        \draw[<->, >=Stealth, very thick, double, double distance=0.5pt] (ward) -- node[below,yshift=-1cm] {\textbf{\small{$\bm{\langle S\rangle_{soft}}$}}} (soft);
        \draw[->, >=Stealth, very thick, dashed] (leaky.south) to [out=-90, in=-230] node[left,pos=0.3,xshift=-1cm] {\textbf{Same Physics}} (soft.north);
    \end{tikzpicture}
    \caption{\small \textbf{The Leaky Correspondence.} The geometry of the leaky boundary defines an infinite-dimensional symmetry (BMS). The conservation of its charges (Ward identity) is mathematically identical to the universal law of soft graviton emission in scattering. They describe the same infrared physics: energy and information inevitably `leak' to infinity.}
    \label{fig:leaky_correspondence}
\end{figure}

Thus, the Ward identity \eqref{eq:ward_identity} is the mathematical expression of a single, deep physical fact, that is, \textsc{the leakiness of the boundary, allowing non-zero news $N_{zz}$, is inseparable from the leakiness of scattering through soft emission.} The infinite-dimensional BMS symmetry, born from the geometry of the leaky null boundary, \textit{dictates} the infrared structure of every gravitational scattering event in the bulk, Fig.\ref{fig:leaky_correspondence}.

\subsection{Organizing the Celestial CFT: Collinear Limits, OPEs, and the \(w_{1+\infty}\) Algebra}

The power of a 2D conformal field theory lies in its ability to organize an infinite amount of data through simple principles, namely the so-called \textit{operator product expansions (OPEs)} and \textit{symmetry algebras}. Celestial holography inherits this organizational power. Remarkably, well-known properties of scattering amplitudes translate into these CFT structures, revealing a hidden and potentially enormous symmetry algebra.

\subsubsection{Collinear Limits and the Celestial OPE}
In standard momentum-space physics, when two external massless particles become collinear, that is, their four-momenta are parallel, a scattering amplitude factorizes in a universal way. This \textbf{collinear limit} is a fundamental property of gauge and gravity amplitudes.

In celestial holography, this limit takes on a new life. When two external particles become collinear, their respective insertion points on the celestial sphere, \(z_i\) and \(z_j\), approach each other. In a 2D CFT, the behavior of operators as their insertion points converge is governed by the \textbf{Operator Product Expansion (OPE)}. Therefore, the collinear factorization of amplitudes must be encoded in the OPE of the corresponding celestial operators \cite{Fan2021, Guevara2021}.

For example, the leading OPE for two positive-helicity gluon operators in CCFT takes a remarkably simple form,
\begin{equation}
\mathcal{O}_{\Delta_1, +}(z_1, \bar{z}_1) \mathcal{O}_{\Delta_2, +}(z_2, \bar{z}_2) \propto -\frac{B(\Delta_1-1, \Delta_2-1)}{z_{12}}\; \mathcal{O}_{\Delta_1+\Delta_2-1, +}(z_2, \bar{z}_2) + \dots,
\label{eq:celestial_ope}
\end{equation}
where \(z_{12} = z_1 - z_2\) and \(B\) is the \textit{Euler Beta function} arising from the Mellin transform. This is derived by taking the collinear limit of the corresponding momentum-space amplitude and then performing the Mellin transform. The OPE coefficient is universal. This directly links a dynamical property of 4D scattering to a kinematic algebraic structure in the 2D CFT, the OPE.

\subsubsection{The Tower of Symmetries: \(w_{1+\infty}\) and Beyond}
The OPE structure hints at a symmetry much larger than the global conformal group. Investigations into the soft symmetries of gauge theory and gravity revealed an infinite tower of conserved charges. These charges are associated not just with leading (\( \omega^{-1} \)) and subleading (\( \omega^{0} \)) soft theorems, as we saw earlier, but with an entire infinite series of \textbf{sub-subleading soft theorems} (\( \omega^{+1}, \omega^{+2}, \dots \)) and correspond to negative integer values of the conformal dimension $\Delta=-1,-2,\dots-\infty$

Remarkably, the algebra of these soft charges acting on the celestial sphere closes into an infinite-dimensional symmetry known as the \textbf{\(w_{1+\infty}\) algebra}, where $1+\infty$ is an abbr. for $1,0,-1,-2,\dots-\infty$\cite{Strominger2021, Guevara2021}. This algebra is generated by operators \(w^{k}_{m}\), where \(k \geq 0\) and \(m\) is an integer mode index. The index \(k\) is related to the conformal weight/dimension $\Delta$, while \(m\) labels the mode on the sphere. The algebra has a rich structure,
\[
[ w^p_m, w^q_n ] \propto \big[ m(q-1) - n(p-1) \big] \,w^{p+q-2}_{m+n} + \dots
\]
In the celestial context, the operators \(w^{k}_{m}\) are understood as modes of specific \textbf{celestial currents} \(W^k(z)\). These currents are constructed from the soft limits of amplitudes. For instance,
\begin{itemize}
    \item The \(k=0\) current is related to the subleading soft graviton and is identified with the \textbf{celestial stress tensor} \(T(z)\).
    \item The \(k=1\) current is related to the sub-subleading soft graviton.
    \item The entire tower organizes all positive-helicity soft graviton symmetries.
\end{itemize}

The action of these \(w_{1+\infty}\) generators on celestial operators is precisely such that their OPEs, like Eq.\eqref{eq:celestial_ope}, are consistent with the algebra. This means the seemingly complicated OPE structure of the celestial CFT is \textit{governed} by this underlying infinite-dimensional symmetry. The \(w_{1+\infty}\) algebra is thus a candidate for the chiral symmetry algebra of the self-dual sector of gravity, and its study is a major frontier in celestial holography \cite{Strominger2021}.

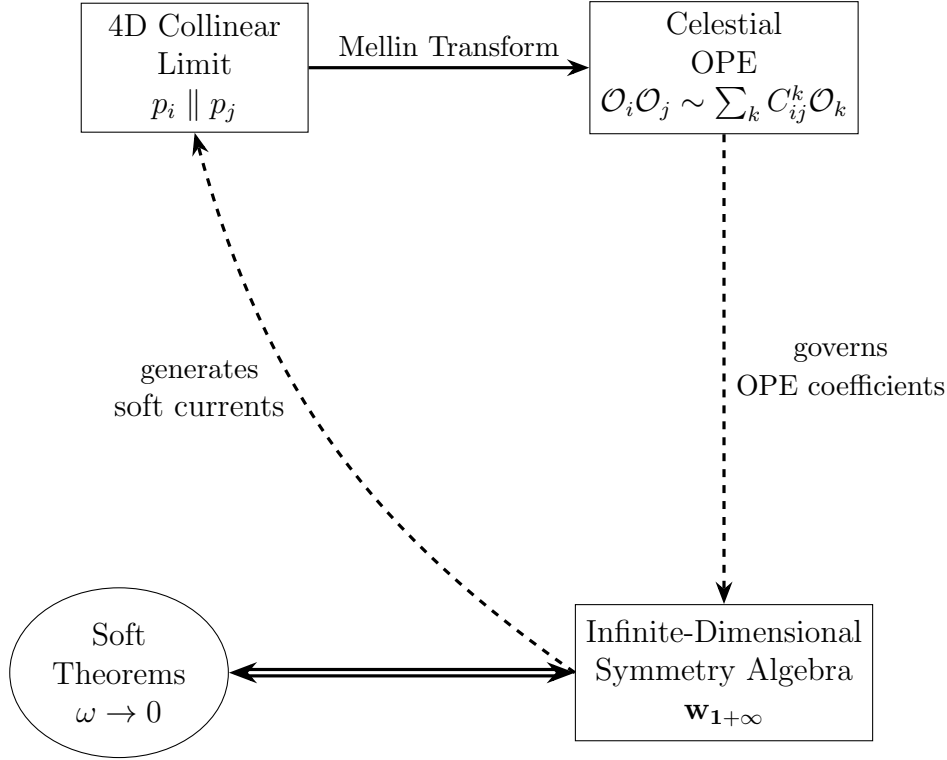
\begin{figure}[h!]
    \centering
    \begin{tikzpicture}[node distance=7cm]
        \node (collinear) [draw, rectangle, minimum width=3cm, minimum height=1.5cm, align=center] {4D Collinear\\Limit\\$p_i \parallel p_j$};
        \node (ope) [draw, rectangle, right of=collinear, minimum width=3cm, minimum height=1.5cm, align=center] {Celestial\\OPE\\$\mathcal{O}_i \mathcal{O}_j \sim \sum_k C_{ij}^k \mathcal{O}_k$};
        \node (walgebra) [draw, rectangle, below of=ope, yshift=-1cm, minimum width=3.5cm, minimum height=1.8cm, align=center] {Infinite-Dimensional\\Symmetry Algebra\\$\mathbf{w_{1+\infty}}$};

        \draw[->, >=Stealth, very thick] (collinear) -- node[above] {\small Mellin Transform} (ope);
        \draw[->, >=Stealth, very thick, dashed] (ope) -- node[right, align=center] {\small governs\\\small OPE coefficients} (walgebra);
        \draw[->, >=Stealth, very thick, dashed, bend left=20] (walgebra.west) to node[left, pos=0.6, align=center] {\small generates\\ soft currents} (collinear.south);

        \node (soft) [draw, ellipse, left of=walgebra, xshift=-1cm, minimum width=2.5cm, align=center] {Soft\\Theorems\\$\omega \to 0$};
        \draw[<->, >=Stealth, very thick, double, double distance=1.2pt] (soft) -- (walgebra);
    \end{tikzpicture}
    \caption{\textbf{From Collinear Limits to Infinite Symmetry.} The collinear limit of 4D scattering amplitudes, when transformed to the celestial basis, defines the operator product expansion (OPE) of the celestial CFT. The structure of these OPEs is controlled by an infinite-dimensional symmetry algebra, \(w_{1+\infty}\), which itself is generated from the tower of soft theorems. This creates a powerful, self-consistent algebraic framework for flat spacetime holography.}
    \label{fig:collinear_ope_w}
\end{figure}

\subsubsection{Physical Interpretation and Implications}
The emergence of the \(w_{1+\infty}\) algebra has astonishing implications,
\begin{itemize}
    \item \textbf{UV/IR Connection:} The algebra mixes transformations that act on hard (high-energy) and soft (low-energy) particles. This is a concrete realization of a UV/IR connection in flat space, where symmetries link the detailed structure of hard scattering to universal infrared phenomena.
    \item \textbf{Constraint on Dynamics:} The symmetry is so constraining that it may \textit{determine} the tree-level S-matrix of certain theories such as self-dual gravity from symmetry principles alone, reminiscent of the bootstrap program in CFT.
    \item \textbf{Carrollian Origin:} There is a natural connection to Carrollian symmetry on \(\mathscr{I}^+\). The \(w_{1+\infty}\) algebra can be understood as originating from the quantization of a particular phase space of radiative modes at null infinity, further solidifying the Carrollian nature of the dual theory.
\end{itemize}

The important fact is that studying collinear limits and OPEs in celestial holography is not merely a translation of known physics into a new language. It is a discovery tool, revealing an intricate and powerful infinite-dimensional symmetry algebra that organizes the S-matrix and offers a new algebraic window into quantum gravity in flat space, Fig\ref{fig:collinear_ope_w}.

\subsection{Mathematical Derivation of the \(w_{1+\infty}\) Algebra}

The infinite-dimensional \(w_{1+\infty}\) algebra is not postulated but derived as the symmetry governing the collinear/soft sector of gravitational scattering. The derivation proceeds in several logical steps, from soft theorems to conserved currents, from currents to their operator product expansion (OPE), and finally from the OPE to the mode algebra.

\paragraph{Step 1: From Soft Theorems to Celestial Currents}
Consider the soft expansion of a positive-helicity graviton operator. In the Mellin basis, taking the soft limit \(\omega \to 0\) corresponds to evaluating the celestial operator \(\mathcal{O}_{\Delta, +2}(z, \bar{z})\) at special integer values of the conformal dimension \(\Delta\). The leading, subleading, and sub-subleading soft graviton theorems are reproduced by the residues of \(\mathcal{O}_{\Delta, +2}\) at \(\Delta = 1, 0, -1, \dots\) \cite{Guevara2021, Strominger2021}.

This motivates the definition of an infinite tower of \textbf{celestial soft currents} as contour integrals in the complex \(\Delta\)-plane,
\begin{equation}
H^k(z) \equiv \oint_{\Delta=1-k} \frac{d\Delta}{2\pi i} \, \mathcal{O}_{\Delta, +2}(z, \bar{z}), \quad k = 0, 1, 2, \dots
\label{eq:soft_currents}
\end{equation}
Here, \(H^0(z)\) is the current associated with the leading soft theorem (supertranslation), \(H^1(z)\) with the subleading (superrotation), and so on. In the language of the Bondi-Sachs metric, these currents \(H^k(z)\) are constructed from specific \(u\)-integrals and derivatives of the shear tensor \(C_{zz}\) and the Bondi news \(N_{zz}\). For instance, \(H^0(z)\) is related to the time-independent part of \(C_{zz}\), while \(H^1(z)\) involves its first \(u\)-moment.

\paragraph{Step 2: The OPE of Soft Graviton Operators}
The fundamental input is the collinear singular behavior of graviton scattering amplitudes. When two positive-helicity gravitons become collinear, their celestial operator product expansion exhibits a universal singularity \cite{Fan2021, Guevara2021}. For two operators \(\mathcal{O}_{\Delta_i, +2}(z_i, \bar{z}_i)\), the singular part of the OPE as \(z_{12} \to 0\) is,
\begin{equation}
\mathcal{O}_{\Delta_1, +2}(z_1, \bar{z}_1) \mathcal{O}_{\Delta_2, +2}(z_2, \bar{z}_2) \sim -\frac{\kappa}{2} \frac{B(\Delta_1-1, \Delta_2-1)}{z_{12}} \sum_{m=0}^\infty \frac{\bar{z}_{12}^m}{m!} {\partial}^m_{\bar{z}} \mathcal{O}_{\Delta_1+\Delta_2-1, +2}(z_2, \bar{z}_2) + \dots,
\label{eq:graviton_ope_detail}
\end{equation}
where \(B(x,y)\) is the \textit{Euler Beta function}, \(\kappa\) is the gravitational coupling, and the sum over \(m\) arises from a Taylor expansion in the $\bar{z}$ coordinate. This OPE encodes the collinear splitting functions of gravitons in celestial language.

\paragraph{Step 3: Mode Expansion and the Algebra}
To extract the algebra, we perform a Laurent expansion of the holomorphic currents \(H^k(z)\),
\begin{equation}
H^k(z) = \sum_{m \in \mathbb{Z}} \frac{w^k_m}{z^{m+k+1}}.
\label{eq:mode_expansion}
\end{equation}
The modes \(w^k_m\) are the generators we seek. Their algebra is determined by the singular part of the OPE between two currents, \(H^{p}(z) H^{q}(w)\). Using the definition \eqref{eq:soft_currents} and the fundamental graviton OPE \eqref{eq:graviton_ope_detail}, one can compute this current-current OPE. After a careful calculation that involves integrating over the conformal dimensions and summing over the anti-holomorphic Taylor series, one finds a compact result \cite{Guevara2021},
\begin{equation}
H^p(z) H^q(w) \propto -\frac{\kappa}{2} \frac{(p+q-2)!}{(p-1)!(q-1)!} \frac{H^{p+q-2}(w)}{(z-w)^2} + \frac{(q-1)H^{p+q-1}(w)}{z-w} +\dots.
\label{eq:current_ope}
\end{equation}
The first term is a double pole, characteristic of a classical central term, while the second term is a single pole indicating the structure of the Lie bracket.

\paragraph{Step 4: Deriving the \(w_{1+\infty}\) Commutators}
Finally, the algebra of the modes \(w^k_m\) is obtained by contour integration of the OPE \eqref{eq:current_ope}. The standard CFT formula for the commutator of modes is,
\begin{equation}
    [w^p_m, w^q_n] = \oint_0 \frac{dz}{2\pi i} \oint_z \frac{dw}{2\pi i} \, z^{m+p} w^{n+q} \, H^p(z) H^q(w).
\end{equation}

Performing this double contour integral on the right-hand side of \eqref{eq:current_ope} yields the celebrated \(w_{1+\infty}\) commutation relations,
\begin{equation}
[w^p_m, w^q_n] = \left[ m(q-1) - n(p-1) \right] w^{p+q-2}_{m+n} + \dots.
\label{eq:w_infinity_commutator}
\end{equation}
This algebra is infinite-dimensional, with the integer \(p \geq 1\) labeling the generator is related to the soft theorem order and \(m \in \mathbb{Z}\) labeling its holomorphic mode.

\subsubsection{Interpretation and Significance}
The derivation establishes several key insights about this algebra,
\begin{itemize}
    \item The generators \(w^k_m\) are \textit{symmetry generators}. Their action on any celestial operator (via the corresponding OPE) generates the soft and collinear singularities mandated by the soft theorems.
    \item This algebra is a candidate for the symmetry algebra that completely constrains the S-matrix of self-dual gravity, pointing toward a celestial conformal field theory with \(w_{1+\infty}\) symmetry.
    \item The \(w_{1+\infty}\) algebra emerges rigorously from the universal infrared structure of gravitational scattering. It is the mathematical embodiment of the infinite tower of soft theorems, organizing them into a single, powerful symmetry principle on the celestial sphere.
\end{itemize}

\subsubsection{A Stop at Virasoro Algebra}

A point of frequent confusion arises in identifying the Virasoro algebra as a subalgebra within \(w_{1+\infty}\). Our derived commutator,
\begin{equation}
[w^p_m, w^q_n] = \big( m(q-1) - n(p-1) \big) w^{p+q-2}_{m+n},
\end{equation}
appears to vanish for \(p=q=1\), yielding \([w^1_m, w^1_n] = 0\). This seems to contradict the Virasoro algebra. The resolution lies in the physical meaning of the index \(p\). In the celestial context, the index \(p\) is tied to the \textit{conformal weight} \(h\) of the current \(H^p(z)\). More precisely, \(H^p(z)\) has holomorphic weight \(h = p+1\). Therefore, the current \(H^0(z)\) has weight \(h=1\). This is the \textbf{celestial stress-energy tensor} \(T(z) \equiv H^0(z)\). Its modes \(L_m \equiv w^0_m\) generate 2D conformal transformations. While the current \(H^1(z)\) has weight \(h=2\). This is a spin-2 current, but it is \textit{not} the stress tensor. It is associated with the subleading soft graviton and generates extended symmetries.

The correct identification is \(L_m = w^0_m\). To find their algebra, we could return to the OPE of the stress tensor with itself, Eq. \eqref{eq:current_ope}, setting \(p=q=0\):
\begin{equation}
T(z) T(w) = H^0(z) H^0(w) \propto -\frac{\kappa}{2} \frac{(-2)!}{(-1)!(-1)!} \frac{H^{-2}(w)}{(z-w)^2} + \frac{(-1)H^{-1}(w)}{z-w}.
\end{equation}
This expression is at best formal due to factorials of negative numbers. One must then compute the OPE for \(H^0\) directly from first principles. The direct calculation, starting from the soft theorem or from the known transformation properties of the shear \(C_{zz}\), yields the standard stress-tensor OPE,
\begin{equation}
T(z) T(w) \sim \frac{2 T(w)}{(z-w)^2} + \frac{\partial T(w)}{z-w} + \frac{c}{2 (z-w)^4},
\end{equation}
where the central charge \(c\) is zero at tree level in Einstein gravity. Performing the mode expansion \(T(z) = \sum_{m} L_m z^{-m-2}\) and applying contour integration gives the Virasoro algebra,
\begin{equation}
[L_m, L_n] = (m-n) L_{m+n} + \frac{c}{12} m(m^2-1) \delta_{m+n,0}.
\end{equation}
Thus, the Virasoro algebra is indeed contained as the \(p=0\) sector and \(L_m = w^0_m\).

The generators \(w^1_m\) on the other hand form a different set. Their commutator from Eq. \eqref{eq:w_infinity_commutator} is,
\begin{equation}
[w^1_m, w^1_n] = (m\times0 - n\times0) w^{0}_{m+n} = 0.
\end{equation}
This is consistent because \(H^1(z)\) is a weight-2 primary field under the Virasoro algebra generated by \(T(z)=H^0(z)\). For a primary field \(\mathcal{O}_h\) of weight \(h\), the OPE with the stress tensor is \(\displaystyle T(z) \mathcal{O}_h(w) \sim \frac{h \mathcal{O}_h(w)}{(z-w)^2} + \frac{\partial \mathcal{O}_h(w)}{z-w}\). The single-pole term in the \(H^1 H^1\) OPE is proportional to \(H^2\), not to \(H^1\) itself, which is why the modes \(w^1_m\) commute among themselves. They do not themselves form a Virasoro algebra.

\section{Carrollian vs Celestial Holography}
\label{subsec:carrollian-celestial-comparison}

The quest for a holographic description of asymptotically flat spacetimes has produced two distinct but deeply related frameworks. \textit{Carrollian holography} and Celestial holography approach the problem from complementary geometric perspectives. Both frameworks share the ultimate goal of encoding quantum gravity in Minkowski space within lower-dimensional theories, yet they employ fundamentally different mathematical structures and physical interpretations. This section provides a comparison of these two approaches.

\subsection{Geometric Foundations and Base Manifolds}

Carrollian holography is fundamentally grounded in the geometry of null infinity. The theory lives on $\mathscr{I}^\pm$. This manifold possesses an intrinsic Carrollian structure characterized by a degenerate metric. The metric on $\mathscr{I}^\pm$ takes the form $ds^2 = q_{AB}dx^Adx^B$ where $q_{AB}$ is a rank-2 tensor on the spherical cross-sections. The null direction along $\mathscr{I}^\pm$ is generated by the vector field $\partial_u$, with $u$ being the retarded time coordinate as before. Crucially, the Lie derivative of the metric along this null direction vanishes: $\mathcal{L}_{\partial_u} q_{AB} = 0$. This structure defines a Carrollian manifold, \textit{which is essentially a fiber bundle with null fibers over the celestial sphere}.

Celestial holography takes a different geometric starting point. This framework works directly on the celestial sphere $\mathbb{S}^2$ obtained by stereographic projection of the asymptotic directions. The celestial sphere is equipped with a standard conformal structure. Complex coordinates $(z,\bar{z})$ parameterize this sphere through the stereographic projection map. The natural metric on the celestial sphere is the round metric $dzd\bar{z}/(1+z\bar{z})^2$, though the conformal structure is what matters most. Unlike Carrollian holography which retains the null direction, celestial holography compactifies the null direction through a Mellin transform. This compactification trades the null coordinate $u$ for a continuous conformal dimension $\Delta$.

The Carrollian framework maintains explicit dependence on retarded time. Physical observables in Carrollian holography are functions or fields on $\mathscr{I}^\pm$, depending on both $u$ and the angular coordinates. The celestial framework, in contrast, works with operators that depend only on the celestial coordinates. Time dependence is encoded in the scaling dimension through the Mellin transform. This fundamental difference in geometric perspective leads to distinct mathematical formulations and physical interpretations.

\subsection{Symmetry Algebras and Their Realizations}

Both frameworks realize the BMS group as their fundamental symmetry algebra, but they implement this symmetry in markedly different ways. Carrollian holography represents BMS symmetries as vector fields acting directly on $\mathscr{I}^\pm$. Supertranslations appear as shifts in the retarded time coordinate: $u \to u + f(z,\bar{z})$. Superrotations are represented as conformal Killing vectors on the celestial sphere. The full BMS algebra appears in its original geometric formulation: $\mathfrak{bms}_4 = \text{Vect}(S^2) \ltimes C^\infty(S^2)$. This realization is geometrically transparent and maintains close contact with the asymptotic structure of spacetime.

Celestial holography goes one step further. It implements BMS symmetries through their action on celestial operators. Supertranslations are represented by operators that insert soft gravitons. Superrotations generate Virasoro transformations on the celestial sphere. The symmetry algebra manifests through the transformation properties of celestial operators $\mathcal{O}_{\Delta,J}(z,\bar{z})$. Under a conformal transformation $z \to z'(z)$, these operators transform as primary fields with weights $(h,\bar{h})$. The celestial framework naturally incorporates infinite-dimensional extensions of the Lorentz group, including full Virasoro and $w_{1+\infty}$ algebras. This algebraic approach enables the application of powerful conformal field theory techniques.

The Carrollian symmetry group extends beyond BMS to include Carrollian boosts. These are transformations that mix the null direction with spatial directions on the sphere. The full Carroll group contains the BMS group as a subgroup but includes additional transformations specific to Carrollian geometry. Celestial holography, in contrast, focuses on the conformal group $SL(2,\mathbb{C})$ and its extensions. The celestial perspective naturally emphasizes the conformal properties of scattering amplitudes when rewritten in a conformal basis.

\subsection{Holographic Dictionaries and Mappings}

The holographic dictionary differs substantially between the two frameworks. Carrollian holography establishes a direct map between bulk gravitational data and boundary fields on $\mathscr{I}^\pm$. The asymptotic expansion of the metric provides boundary data: the Bondi mass aspect $m_B$, angular momentum aspect $N_A$, and shear tensor $C_{AB}$. The News tensor $N_{AB} = \partial_u C_{AB}$ encodes gravitational radiation flux through null infinity. Memory effects are naturally captured by changes in the shear tensor: $\Delta C_{AB} = \int du N_{AB}$. This geometric dictionary maintains clear physical interpretation in terms of measurable gravitational observables.

Celestial holography employs a transformative dictionary based on scattering amplitudes. The fundamental map takes momentum-space scattering amplitudes and rewrites them in a conformal basis through a Mellin transform. Specifically, celestial operators are defined as $\mathcal{O}_{\Delta,J}(z,\bar{z}) = \int_0^\infty d\omega \omega^{\Delta-1} a(\omega\hat{q}(z,\bar{z}), J)$. This transformation converts scattering amplitudes into correlation functions on the celestial sphere: $\langle \mathcal{O}_{\Delta_1,J_1}(z_1,\bar{z}_1) \cdots \mathcal{O}_{\Delta_n,J_n}(z_n,\bar{z}_n) \rangle$. The conformal weights are related to the energy scaling of amplitudes, with $\Delta = 1 + i\lambda$ for massless particles lying in the principal series representation.

Carrollian currents emerge naturally from the asymptotic symmetry analysis. These currents $J_u$ and $J_A$ are constructed from the boundary data and generate the asymptotic symmetries through Poisson brackets. Celestial currents, in contrast, arise from soft theorems. The soft graviton theorem translates into a Ward identity for supertranslation symmetry. Subleading soft theorems correspond to Ward identities for superrotation symmetry. This connection between soft theorems and asymptotic symmetries provides a powerful consistency check for both frameworks.

\subsection{Mathematical Structures and Technical Tools}

Carrollian holography employs mathematical tools adapted to null hypersurfaces. The fundamental functional space is $\mathcal{F}(\mathscr{I}^+)$, consisting of smooth functions or fields on null infinity. Quantization proceeds through canonical methods adapted to null surfaces, with the retarded time $u$ playing the role of time parameter. Propagators are Carrollian Green functions that satisfy equations adapted to the Carrollian structure. The Carrollian limit of field theories is obtained by taking the speed of light to zero, which suppresses certain kinetic terms. This limit produces distinctive equations of motion with only first-order time derivatives.

Celestial holography leverages the full machinery of two-dimensional conformal field theory. The state space $\mathcal{H}_{\text{celestial}}$ is organized into representations of $SL(2,\mathbb{C})$. Radial quantization on the celestial sphere provides a powerful organizational principle. Operator product expansions become central tools, with celestial OPEs encoding collinear limits of scattering amplitudes. Conformal blocks control the representation theory of the celestial symmetry algebra. The bootstrap program offers a systematic approach to constraining celestial correlators through symmetry, analyticity, and crossing symmetry.

Carrollian field theories exhibit distinctive features due to their degenerate geometry. The Carrollian action for a scalar field takes the form $S_{\text{Carroll}} = \int du d^2z \sqrt{q} \left[ i\Phi^\dagger \partial_u \Phi + \kappa \partial_A \Phi^\dagger \partial^A \Phi \right]$. Notice the absence of $(\partial_u)^2$ terms, which reflects the Carrollian causality structure. Celestial conformal field theories, in contrast, resemble standard two-dimensional CFTs but with continuous spin and unusual operator dimensions. The celestial OPE takes the form $\mathcal{O}_{\Delta_1,J_1}(z,\bar{z})\mathcal{O}_{\Delta_2,J_2}(0,0) \sim \sum_k C_{12}^k z^{h_k-h_1-h_2} \bar{z}^{\bar{h}_k-\bar{h}_1-\bar{h}_2} \mathcal{O}_{\Delta_k,J_k}(0,0)$.

\subsection{Physical Interpretations and Domain of Applicability}

Carrollian holography offers several compelling physical interpretations. The framework directly describes gravitational radiation as measured by asymptotic observers. Memory effects are naturally encoded in the boundary metric components. The connection to fluid dynamics emerges through the Carrollian limit of relativistic hydrodynamics. Black hole physics finds a natural home in Carrollian holography, with the event horizon dynamics potentially related to Carrollian dynamics at $\mathscr{I}^+$. The framework maintains geometric transparency throughout, with clear connections to gravitational observables.

Celestial holography provides powerful insights into scattering amplitudes. The framework recasts flat space scattering as a two-dimensional conformal field theory problem. This perspective reveals hidden infinite-dimensional symmetries in gauge and gravitational theories. The celestial OPE provides a novel organizing principle for collinear and soft limits. Massive particles can be incorporated through different representations of $SL(2,\mathbb{C})$, offering potential unification of massless and massive sectors. Connections to AdS/CFT emerge through analytic continuation and flat space limits.

Both frameworks address the infrared structure of quantum gravity. Carrollian holography naturally encodes infrared divergences through integrals over retarded time. Celestial holography captures soft and collinear singularities through poles in the conformal dimension plane. The two approaches offer complementary insights into the infrared structure of scattering amplitudes and its connection to asymptotic symmetries.

\subsection{Key Results and Open Challenges}

Carrollian holography has established several important results. The Carrollian limit of Einstein's equations yields constraint equations on $\mathscr{I}^+$ that govern the evolution of boundary data. Carrollian energy-momentum tensors have been constructed from subleading terms in the asymptotic expansion. Connections to the fluid/gravity correspondence emerge in the ultra-relativistic limit. However, significant challenges remain. A non-perturbative formulation of Carrollian holography is still lacking. The renormalization of Carrollian field theories presents novel difficulties due to their unusual causal structure. Interacting Carrollian quantum field theories are not well understood, particularly at strong coupling.

Celestial holography has produced compelling evidence for its consistency. Celestial amplitudes exhibit conformal covariance under $SL(2,\mathbb{C})$ transformations. Soft theorems have been shown to be equivalent to Ward identities for BMS symmetries. Explicit celestial OPEs have been computed for gauge theory and gravity, revealing their algebraic structure. Mounting evidence points to a $w_{1+\infty}$ symmetry governing the self-dual sector of gravity. Yet open questions persist. A Lagrangian formulation of celestial CFT remains elusive. The classification of possible celestial conformal field theories is incomplete. The problem of bulk reconstruction from celestial data is largely unsolved. The proper treatment of massive particles and black holes requires further development.

\subsection{Bridging the Two Frameworks}

Recent developments suggest that Carrollian and celestial holography are not competing paradigms but complementary perspectives. A precise dictionary connects the two frameworks through integral transforms. Carrollian fields on $\mathscr{I}^+$ can be expressed as inverse Mellin transforms of celestial operators,

\begin{equation}
    \Phi(u,z,\bar{z}) = \int \frac{d\Delta}{2\pi i} u^{-\Delta} \mathcal{O}_{\Delta,J}(z,\bar{z}). 
\end{equation}
This relation establishes that the two frameworks are related by a change of basis, much like position and momentum space in quantum mechanics.

The celestial OPE finds a natural interpretation in Carrollian language. The OPE singularity $z_{12} \to 0$ corresponds to the collinear limit of massless particles. In Carrollian terms, this limit describes wavefronts on $\mathscr{I}^+$ that become coincident along null generators. The coefficients in the celestial OPE encode universal behavior of scattering amplitudes in collinear limits. These same coefficients constrain the operator algebra of Carrollian fields through their transformation properties under asymptotic symmetries.

Both frameworks share common challenges that point toward deeper structure. Understanding the holographic encoding of massive particles remains problematic in both approaches. Black hole formation and evaporation present puzzles for both Carrollian and celestial descriptions. The complete holographic encoding of bulk information, particularly behind horizons, requires insights from both geometric and algebraic perspectives.

The Carrollian approach excels in geometric transparency and direct connection to gravitational observables. The celestial approach provides powerful algebraic methods and connections to amplitude physics. A complete theory of flat space holography will likely synthesize insights from both frameworks. The geometric intuition of Carrollian holography combined with the algebraic power of celestial holography may ultimately yield a comprehensive description of quantum gravity in asymptotically flat spacetimes.

\section{Beyond the Horizon: Current Frontiers and Open Questions}

Celestial holography is a vibrant and rapidly developing field. Its very novelty means it is defined by a set of compelling open questions that drive current research.

\subsection{The Nature of the Celestial CFT}
\begin{itemize}
    \item[Q1] \textbf{What are the fundamental axioms?} We have a proposed dictionary for correlators, but what are the underlying principles (beyond symmetries) that define the CCFT itself? Is it a standard unitary CFT, or something more exotic like a Logarithmic CFT or a Carrollian CFT in disguise?
    \item[Q2] \textbf{What is the celestial state dual to a black hole?} If the CCFT describes flat space quantum gravity, it must contain states corresponding to black holes. How are black hole entropy, evaporation, and information recovery encoded in the celestial correlators \cite{Hawking2015}?
    \item[Q3] \textbf{Can we bootstrap it?} Can we use the conformal bootstrap program by constraining CFTs using symmetries, crossing symmetry, and unitarity to derive the properties of the CCFT from first principles?
\end{itemize}

\subsection{Mathematical Structure and Amplitudes}
\begin{itemize}
    \item[Q4] \textbf{What is the full symmetry algebra?} BMS is a key part, but recent work suggests even larger symmetry structures, the $w_{1+\infty}$ organizing algebra is present in the theory. How do these organize the CCFT spectrum \cite{Chang2023}?
    \item[Q5] \textbf{How do we compute efficiently?} The Mellin transform integrals can be technically challenging. Developing new methods, like the \textit{split representation} for celestial amplitudes \cite{Chang2023}, is an active area of research to compute higher-point and loop-level correlators.
    \item[Q6] \textbf{Massive particles?} The formalism is clearest for massless external states. How do we consistently include massive particles, whose momenta do not map directly to a point on the celestial sphere?
\end{itemize}

\subsection{Connections to Reality and Other Approaches}
\begin{itemize}
    \item[Q7] \textbf{Is there any experimental signature?} This is the long-term dream. Could subtle patterns in gravitational wave memory effects or in the infrared structure of particle collider data be interpreted as signatures of the underlying celestial/Carrollian symmetry?
    \item[Q8] \textbf{What is the precise AdS/CFT $\to$ Flat Space limit?} Understanding how to carefully take the radius of AdS to infinity and recover flat space celestial correlators from AdS boundary correlators is a crucial consistency check and a source of technical insight.
\end{itemize}

\subsection{Soft Hair and Black Hole Microstates}
This connection leapt into the spotlight with the proposal of black hole \textit{soft hair} \cite{Hawking2015}. The idea is that black holes can carry an infinite set of BMS charges, in particular, those of supertranslation, on their horizons, providing a potential reservoir for quantum information. While a classical Schwarzschild black hole is uniquely defined by its \textit{mass}, \textit{charge}, and \textit{angular momentum} as stated by the \textit{no-hair theorem}, these BMS charges represent an infinite family of \textit{degenerate} classical solutions with the same classical attributes. In the quantum theory, these could label microstates, offering a mechanism to store information and potentially address the black hole information paradox.

\section{Bon Voyage!}

Our whirlwind tour has taken us from the comfortable, reflective walls of AdS to the leaky, dynamical frontier of flat space null infinity. We've seen how the Bondi-Sachs formalism \cite{Madler2016} provides the geometry, how the infinite-dimensional BMS symmetry \cite{Madler2016} governs the physics of soft radiation and black hole hair \cite{Hawking2015}, and how the Mellin transform provides a concrete map from 4D scattering to 2D celestial correlators.

Celestial holography offers a powerful framework. It starts from the bedrock of well-tested physics, asymptotic symmetries, soft theorems, and the S-matrix, and attempts to build a holographic description of quantum gravity in our universe from there. It connects deep questions about black holes \cite{Hawking2015}, the nature of the gravitational S-matrix, and the fundamentals of quantum field theory in a new and unifying language.

The journey is just beginning. The celestial sphere, that ancient map of the heavens on which so much dreams and hopes are printed, may yet prove to be the screen on which the quantum hologram of our spacetime is projected.

\section{Acknowledgment}
I would like to thank \textit{Prof. Laura Donnay} for the superb lectures on Celestial Holography she gave during the \textit{Basics of Quantum Gravity school (January 2026)} organized by the International Society for Quantum Gravity (ISQG). These lectures were a perfect combination of subject high attractiveness and excellent teaching skills. A special thank to \textit{Dr. Alicia Castro}, \textit{Dr. Iva Lovrekovic}, \textit{Dr. V H Satheeshkumar} and \textit{Dr. Renata Ferrero} for the work they do to make these crucial educational opportunities available.
\bibliographystyle{unsrt}
\bibliography{references}

@article{Maldacena1999,
    author = {Maldacena, Juan},
    title = {The Large-N Limit of Superconformal Field Theories and Supergravity},
    journal = {International Journal of Theoretical Physics},
    volume = {38},
    number = {4},
    pages = {1113--1133},
    year = {1999},
    doi = {10.1023/A:1026654312961},
    note = {The seminal paper proposing the AdS/CFT correspondence, the foundational idea behind holography.}
}

@article{Madler2016,
    author = {Mädler, Thomas and Winicour, Jeffrey},
    title = {Bondi-Sachs Formalism},
    journal = {Scholarpedia},
    volume = {11},
    number = {12},
    pages = {33528},
    year = {2016},
    doi = {10.4249/scholarpedia.33528},
    note = {An excellent, authoritative review of the Bondi-Sachs formalism, which introduces null infinity $\mathscr{I}^+$ and the BMS group.}
}

@article{Strominger2017,
    author = {Strominger, Andrew},
    title = {Lectures on the Infrared Structure of Gravity and Gauge Theory},
    journal = {arXiv preprint},
    volume = {arXiv:1703.05448},
    year = {2017},
    note = {A comprehensive set of lectures detailing the deep connections between asymptotic symmetries (BMS), soft theorems, and gravitational memory.}
}

@misc{Hawking2015,
    author = {Hawking, Stephen W. and Perry, Malcolm J. and Strominger, Andrew},
    title = {Soft Hair on Black Holes},
    year = {2015},
    eprint = {1601.00921},
    archivePrefix = {arXiv},
    primaryClass = {hep-th},
    note = {The influential paper proposing that black holes carry BMS "soft hair" as a potential reservoir for quantum information.}
}

@article{Chang2023,
    author = {Chang, Chi-Ming and Huang, Yu-tin and Ma, Wen-Jie},
    title = {Split representation in celestial holography},
    journal = {arXiv preprint},
    volume = {arXiv:2311.08736},
    year = {2023},
    note = {An example of current research developing new computational techniques (split representations) for celestial amplitudes.}
}

@article{Pasterski2021,
    author = {Pasterski, Sabrina},
    title = {Lectures on Celestial Amplitudes},
    journal = {The European Physical Journal C},
    volume = {81},
    number = {12},
    pages = {1062},
    year = {2021},
    doi = {10.1140/epjc/s10052-021-09748-8},
    note = {A clear and pedagogical introduction to the core concepts of celestial amplitudes and the Mellin transform.}
}

@article{Donnay2022,
    author = {Donnay, Laura and Puhm, Andrea},
    title = {Celestial holography: an asymptotic symmetry perspective},
    journal = {arXiv preprint},
    volume = {arXiv:2212.09777},
    year = {2022},
    note = {A comprehensive modern review of the field from the perspective of asymptotic symmetries.}
}

@article{Esposito2024,
    author = {Esposito, Giampiero and Vitale, Giuseppe Filiberto},
    title = {Homogeneous Projective Coordinates for the Bondi-Metzner-Sachs Group},
    year = {2024},
    eprint = {2406.00419},
    archivePrefix = {arXiv},
    primaryClass = {gr-qc},
    note = {A detailed study of the BMS group in projective coordinates, discussing the geometry of cuts of null infinity and the action of BMS transformations.}
}

@article{Penrose1967,
    author = {Penrose, Roger},
    title = {Twistor Algebra},
    journal = {Journal of Mathematical Physics},
    volume = {8},
    number = {2},
    pages = {345--366},
    year = {1967},
    doi = {10.1063/1.1705200},
    note = {The seminal paper introducing twistor theory, proposing a fundamental framework where spacetime points are derived from more primitive twistor objects.}
}

@misc{WikipediaTwistor,
    author = {{Wikipedia contributors}},
    title = {Twistor Theory},
    year = {2024},
    url = {https://en.wikipedia.org/wiki/Twistor_theory},
    note = {A comprehensive overview of twistor theory, its history, the twistor correspondence, and its applications in physics.}
}

@misc{TwistorLi,
    author = {Various},
    title = {Twistor Theory},
    year = {2024},
    url = {https://twistor.li/},
    note = {A dedicated resource on twistor theory, covering its mathematical foundations, the Penrose transform, and applications to modern physics.}
}

@article{Mason2011,
    author = {Mason, L. J. and Skinner, D.},
    title = {Conformal Field Theories in Six-Dimensional Twistor Space},
    journal = {Journal of Geometry and Physics},
    volume = {62},
    number = {12},
    pages = {2353--2375},
    year = {2012},
    doi = {10.1016/j.geomphys.2012.08.001},
    eprint = {1111.2585},
    archivePrefix = {arXiv},
    primaryClass = {hep-th},
    note = {Explores the relationship between conformal field theories in twistor space and spacetime, relevant for understanding the twistor formulation of scattering amplitudes.}
}

@article{Donnay2023,
    author = {Donnay, Laura},
    title = {Celestial holography: an asymptotic symmetry perspective},
    journal = {Physics Reports},
    volume = {1033},
    pages = {1--79},
    year = {2023},
    issn = {0370-1573},
    doi = {10.1016/j.physrep.2023.09.004},
    note = {A modern and comprehensive review of the field from the perspective of asymptotic symmetries (BMS group) and soft theorems. It is an invited review for Physics Reports.},
    eprint = {2310.12922},
    archivePrefix = {arXiv},
    primaryClass = {hep-th}
}

@article{Raclariu2021,
  journal={},
   author = {Ana-Maria Raclariu},
   month = {7},
   title = {Lectures on Celestial Holography},
note = {These notes consist of lectures on celestial holography given at the Pre-Strings school 2021.},
   url = {https://arxiv.org/pdf/2107.02075},
   year = {2021}
}

@article{Fan2021,
    author = {Fan, Weizhen and Fotopoulos, Angelos and Taylor, Tomasz R.},
    title = {Soft Limits of Yang-Mills Amplitudes and Conformal Correlators},
    journal = {J. High Energ. Phys.},
    volume = {2021},
    number = {5},
    pages = {203},
    year = {2021},
    doi = {10.1007/JHEP05(2021)203},
    eprint = {1903.01676},
    archivePrefix = {arXiv},
    primaryClass = {hep-th},
    note = {Derives the celestial OPE from collinear limits of gluon amplitudes, establishing a key dictionary entry.}
}

@article{Guevara2021,
    author = {Guevara, Alfredo and Himwich, Elizabeth and Pate, Monica and Strominger, Andrew},
    title = {Holographic symmetry algebras for gauge theory and gravity},
    journal = {J. High Energ. Phys.},
    volume = {2021},
    number = {11},
    pages = {152},
    year = {2021},
    doi = {10.1007/JHEP11(2021)152},
    eprint = {2103.03961},
    archivePrefix = {arXiv},
    primaryClass = {hep-th},
    note = {A foundational paper identifying the $w_{1+\infty}$ algebra as the symmetry governing celestial OPEs in gravity and gauge theory.}
}

@article{Strominger2021,
    author = {Strominger, Andrew},
    title = {$w_{1+\infty}$ and the Celestial Sphere},
    year = {2021},
    eprint = {2105.14346},
    archivePrefix = {arXiv},
    primaryClass = {hep-th},
    journal = {arXiv preprint},
    note = {Explores the $w_{1+\infty}$ algebra as an infinite-dimensional symmetry of the celestial sphere and its connection to soft theorems.}
}

\end{document}